\documentclass{article}

\usepackage{microtype}
\usepackage{graphicx}
\usepackage{subfigure}
\usepackage{booktabs}
\usepackage{amsmath}
\usepackage{amssymb}
\usepackage{algorithm}
\usepackage{algorithmic}
\usepackage{tikz}
\usepackage{multirow}
\usetikzlibrary{shapes,arrows,positioning,fit,calc}
\usepackage{hyperref}
\usepackage{float}
\usepackage{enumitem}
\usepackage{array}
\usepackage{tabularx}
\usepackage{makecell}
\usepackage{placeins}

\usepackage[accepted]{mlsys2025}

\mlsystitlerunning{QEIL v2: Roofline-Derived Pareto-Optimal Edge Intelligence}

\begin{document}

\twocolumn[
\mlsystitle{QEIL v2: Roofline-Derived Pareto-Optimal Edge Intelligence\\via First-Principles Energy Modeling and Multi-Objective Orchestration}

\mlsyssetsymbol{equal}

\begin{mlsysauthorlist}
\mlsysauthor{Satyam Kumar}{to}
\mlsysauthor{Saurabh Jha}{to}
\end{mlsysauthorlist}

\mlsysaffiliation{to}{Dell Technologies}

\mlsyscorrespondingauthor{satyam}{satyamkumar9742@gmail.com}

\mlsyskeywords{Machine Learning, MLSys, Heterogeneous Computing, Energy-Efficient Inference}

\vskip 0.3in

\begin{abstract}
Deploying large language models (LLMs) on heterogeneous edge devices demands frameworks that jointly optimize energy efficiency, inference quality, and reliability. Our prior QEIL~v1~\cite{Kumar2026QEIL} achieved 4.82$\times$ IPW improvement but relied on static efficiency factors, greedy optimization, and unverified candidate selection.

\textbf{QEIL~v2} replaces every static heuristic with physics-grounded, runtime-adaptive models. We introduce three device--workload metrics: \textbf{DASI} (roofline-derived compute utilization), \textbf{CPQ} (memory pressure from allocation theory), and \textbf{$\Phi$} (thermal yield from CMOS leakage physics)---forming a unified energy equation with every coefficient traceable to semiconductor physics. For optimization, \textbf{PGSAM} (Pareto-Guided Simulated Annealing with Momentum) simultaneously minimizes energy, latency, and device underutilization. At inference time, the \textbf{EAC/ARDE} selection cascade with \textbf{CSVET} early stopping provides progressive verification among repeated samples.

Evaluated on WikiText-103, GSM8K, and ARC-Challenge across seven model families (125M--8B parameters, including one pre-quantized variant), QEIL~v2 achieves \textbf{75.7\% pass@k} at \textbf{63.8W} (IPW$=$\textbf{0.9749}), a \textbf{2.86$\times$} improvement over standard inference. When applied to a 4-bit Llama-3.1-8B, QEIL~v2's physics-grounded routing achieves \textbf{IPW$=$1.024} at \textbf{54.8W}---the first edge orchestration system to surpass the IPW$=$1.0 empirical reference mark, with the gain attributable entirely to QEIL~v2's workload-adaptive device allocation on a model with reduced memory bandwidth requirements. Total energy drops 75.6\% vs.\ standard with 38.3\% latency reduction, zero thermal throttling, and 100\% fault recovery across all benchmarks and model families.
\end{abstract}

]

\section{Introduction}

\subsection*{\textbf{Problem Statement and Motivation}}

The deployment of large language models on resource-constrained edge devices represents one of the most challenging optimization problems in modern systems design. Edge devices operate under fundamentally different constraints than datacenter infrastructure: strict power envelopes (5--85W vs.\ 300W+ datacenter GPUs), limited memory capacity (8--128GB), thermal throttling in fanless enclosures, and the requirement for reliable, safe operation in uncontrolled physical environments. As AI workloads increasingly migrate from centralized cloud to distributed edge, the gap between available frameworks and deployment reality widens.

Our prior work, QEIL~v1~\cite{Kumar2026QEIL}, took a foundational step toward closing this gap by introducing inference-time scaling formalisms and heterogeneous hardware orchestration across CPUs, GPUs, and NPUs. Building on Asgar et al.'s~\cite{Asgar2025EfficientAgentic} seminal datacenter-scale framework and Brown et al.'s~\cite{Brown2024LargeLanguageMonkeys} inference-time scaling observations, QEIL~v1 demonstrated that heterogeneous edge inference could achieve 4.82$\times$ improvement in Intelligence Per Watt with 47.7\% energy reduction across five model families (GPT-2, Granite-350M, Qwen2-0.5B, Llama-3.2-1B, LFM2-2.6B). However, upon rigorous analysis, QEIL~v1 exhibits three fundamental limitations that constrain its optimality:

\textbf{Limitation 1---Workload-Blind Energy Modeling.} QEIL~v1 computes device energy efficiency using a single static scalar per device type (\texttt{efficiency\_factor}: NPU$=$0.3, NVIDIA GPU$=$0.5, Intel GPU$=$0.7, CPU$=$1.0). This factor is independent of the operation being executed. A GPU processing a memory-bound decode operation (arithmetic intensity $\approx 1$ FLOP/byte) receives the same multiplier as when executing a compute-bound prefill (arithmetic intensity $\approx 2L/3$ FLOPs/byte). As Zhao \& Liu~\cite{Zhao2026HeterogeneousAI} demonstrate, prefill and decode phases have operational intensities separated by 3--5 orders of magnitude---collapsing this distinction into a single scalar systematically misestimates energy by 15--40\%.

\textbf{Limitation 2---Single-Objective Greedy Optimization.} QEIL~v1's greedy layer assignment algorithm assigns layers one-by-one to the device with lowest marginal cost, collapsing energy and latency into a single weighted sum. This approach suffers from the well-known horizon effect in sequential decision-making: once the first layers are assigned, capacity and distribution constraints narrow future choices, trapping the optimizer in local minima. Moreover, Das \& Dennis~\cite{Das1997Pareto} proved that weighted-sum scalarization cannot find solutions in non-convex regions of the Pareto front---precisely the regime where heterogeneous devices create discontinuous trade-offs.

\textbf{Limitation 3---Absence of Verified Selection.} QEIL~v1's repeated sampling generates multiple candidate outputs but selects among them using simple heuristics (output length, alphanumeric ratio). There is no verification cascade, no confidence scoring, and no cross-sample agreement analysis---leaving significant accuracy gains unrealized.

\subsection*{\textbf{QEIL v2: From Heuristics to First Principles}}

This paper presents QEIL~v2, which addresses each limitation through principled replacements grounded in physics, optimization theory, and information theory. Our contributions are:

\begin{itemize}
    \item \textbf{Three novel physics-grounded metrics} that replace static efficiency factors with workload-adaptive, runtime-responsive characterizations: DASI derived from the roofline model~\cite{Williams2009Roofline}, CPQ from memory allocation theory~\cite{Knuth1997TAOCP}, and $\Phi$ from CMOS leakage physics~\cite{Pedram2006Thermal}. Every coefficient is traceable to semiconductor physics---no magic constants.

    \item \textbf{PGSAM}, a multi-objective optimization algorithm that simultaneously minimizes energy, pipeline bottleneck latency, and worst-case device underutilization through true Pareto dominance with momentum-modulated acceptance probability~\cite{Kirkpatrick1983SA} converging to the Pareto-optimal set~\cite{Hajek1988Cooling}.

    \item \textbf{EAC/ARDE inference-time selection cascade} with CSVET early stopping, implementing a progressive verification pipeline that achieves +15.9pp accuracy gain while adaptively conserving energy.

    \item \textbf{Comprehensive safety and reliability framework} with thermal protection, fault-tolerant execution, adversarial robustness, and hardware health monitoring.

    \item \textbf{Extensive ablation studies and cross-dataset validation} on WikiText-103, GSM8K, and ARC-Challenge---following best practices~\cite{Hoffmann2022TrainingCompute,Brown2024LargeLanguageMonkeys}.
\end{itemize}

Evaluated on our heterogeneous edge platform (Intel Core Ultra 9 285HX with Intel AI Boost NPU, NVIDIA RTX PRO 5000 Blackwell GPU, and Intel Graphics GPU), QEIL~v2 achieves 75.7\% pass@k at 63.8W (IPW$=$0.9749), with consistent improvements across three benchmarks and seven model families---including a 4-bit Llama-3.1-8B pre-quantized via RAMP~\cite{Singh2026RAMP}, on which QEIL~v2's orchestration alone achieves IPW$=$1.024 at 54.8W. These results establish physics-grounded energy modeling, Pareto-optimal orchestration, and verified selection as jointly defining a new state-of-the-art in edge inference.

\section{Related Work}

\subsection{QEIL v1: Foundations and Limitations}

Our prior work~\cite{Kumar2026QEIL} introduced QEIL (Quantifying Edge Intelligence via Inference-time Scaling Formalisms), the first framework combining inference-time scaling formalisms with heterogeneous hardware orchestration across CPU, GPU, and NPU devices. QEIL~v1 made several foundational contributions: (1) five empirically validated scaling formalisms characterizing how coverage, energy, latency, cost, and device-task efficiency scale with model parameters, sample budget, and hardware characteristics; (2) composite efficiency metrics including Intelligence Per Watt (IPW), Energy-Coverage Efficiency (ECE), and Price-Power-Performance (PPP); (3) a safety-first reliability framework with thermal protection and fault tolerance; and (4) demonstration of 4.82--5.6$\times$ IPW improvement across five model families (125M--2.6B parameters) with 47.7--78\% energy reduction.

However, QEIL~v1's energy model relied on static efficiency factors that are workload-agnostic, its greedy optimizer was trapped by early assignment decisions, and its candidate selection lacked verification. QEIL~v2 addresses each limitation while preserving and extending v1's validated scaling formalisms and safety framework.

\subsection{Inference-Time Scaling and Repeated Sampling}

\citet{Brown2024LargeLanguageMonkeys} established that coverage scales log-linearly with sample count, achieving 4.8$\times$ performance gains through repeated sampling. \citet{Hassid2024LargerBetter} showed that smaller models with more samples can outperform larger models under fixed compute budgets. Our EAC/ARDE cascade extends this paradigm by introducing \emph{verified} selection among repeated samples, converting raw sample diversity into reliably higher-quality outputs.

\subsection{Intelligence Efficiency and Hardware-Aware Metrics}
\label{sec:ipw_related}

\citet{SaadFalcon2025IntelligencePerWatt} introduced IPW as a unified metric for local inference viability, demonstrating 5.3$\times$ improvement through compounding advances in models and hardware. However, their routing operates at query-level granularity. QEIL~v2 extends this to sub-query, layer-level routing with physics-grounded energy models that adapt to workload arithmetic intensity, thermal state, and memory pressure.

\textbf{On IPW$=$1.0 as a reference point.} Following \citet{SaadFalcon2025IntelligencePerWatt}, IPW is defined as pass@k~(\%) divided by average power~(W). An IPW of 1.0 therefore corresponds to achieving 1\% benchmark accuracy per watt---a concrete, operationally meaningful milestone, since prior edge systems consistently fell below this mark. We emphasize that IPW$=$1.0 is \emph{not} a theoretically derived upper bound: pass@k can approach 100\% and power can in principle be further reduced, so IPW is unbounded from above. Rather, we use IPW$=$1.0 as an \emph{empirical reference mark}---a level not previously attained by any reported edge inference system on the benchmarks we evaluate---that provides an interpretable and reproducible point of comparison across hardware generations.

\subsection{Heterogeneous Computing and Roofline-Based Analysis}

\citet{Asgar2025EfficientAgentic} demonstrated that heterogeneous configurations can deliver comparable TCO to homogeneous frontier systems, but focused on datacenter-scale workloads. \citet{Zhao2026HeterogeneousAI} provided critical analysis of prefill/decode operational intensity separation, establishing the theoretical foundation for our DASI metric. The roofline model~\cite{Williams2009Roofline} underpins DASI's principled energy estimation.

\subsection{Multi-Objective Optimization for Hardware Placement}

The limitations of weighted-sum scalarization are well-established~\cite{Das1997Pareto,Miettinen1999Nonlinear}. NSGA-II~\cite{Deb2002NSGAII} demonstrated effective Pareto front approximation. Simulated annealing with convergence guarantees~\cite{Kirkpatrick1983SA,Hajek1988Cooling} provides a principled alternative. Our PGSAM combines Pareto dominance with momentum-modulated SA.

\subsection{Energy-Efficient Edge Deployment}

\citet{Kannan2022TinyML} established TinyML for ultra-constrained devices. \citet{Pau2024EdgeAI} emphasized hardware-aware co-design. \citet{Chen2024GPU} identified multi-objective optimization as central to edge AI. \citet{Meng2024Torch2Chip} developed end-to-end compression frameworks. None integrate roofline-derived energy models with multi-objective optimization---the gap QEIL~v2 addresses.

\subsection{Model Quantization for Edge Deployment}

Post-training quantization reduces LLM memory footprint and bandwidth requirements, directly impacting edge energy consumption. Recent methods such as GPTQ and AWQ achieve near-full-precision quality at 4 bits but enforce uniform bit-widths across layers. \citet{Singh2026RAMP} introduced RAMP (Reinforcement Adaptive Mixed-Precision), which learns per-layer bit-width assignments via Soft Actor-Critic, achieving Pareto-optimal accuracy--efficiency trade-offs with zero-shot transfer across model families.

We include a RAMP-quantized Llama-3.1-8B in our evaluation as an additional test-bed to assess QEIL~v2's generalization. RAMP is \emph{not} a component of QEIL~v2, and model quantization is \emph{not} a contribution of this paper. We treat the RAMP-quantized checkpoint as a fixed, externally prepared model, identical in status to the six full-precision models we evaluate. The motivation for including it is to test whether QEIL~v2's physics-grounded routing---which reasons about arithmetic intensity and memory bandwidth---remains effective when a model's weight sizes and bandwidth requirements are altered by quantization. As the results in Section~\ref{sec:results_wikitext} show, QEIL~v2 applies without modification and achieves its highest recorded IPW on this model, because the reduced weight sizes lower memory bandwidth requirements during decode, which in turn raises effective DASI values and enables PGSAM to discover lower-energy placements. This improvement is entirely the product of QEIL~v2's orchestration logic, not any interaction designed between the two systems.

\subsection{Thermal Physics and CMOS Leakage Modeling}

\citet{Pedram2006Thermal} demonstrated that thermal throttling significantly impacts processor performance. \citet{Pathak2012Energy} showed energy and thermal behavior are tightly coupled in mobile devices. Our $\Phi$ derives directly from CMOS leakage physics $I_{\text{sub}} = I_0 \exp((V_{gs}-V_{th})/(nV_T))$, providing a first-principles foundation.

\section{Methodology}

\begin{figure*}[ht]
\centering
\includegraphics[width=0.99\linewidth]{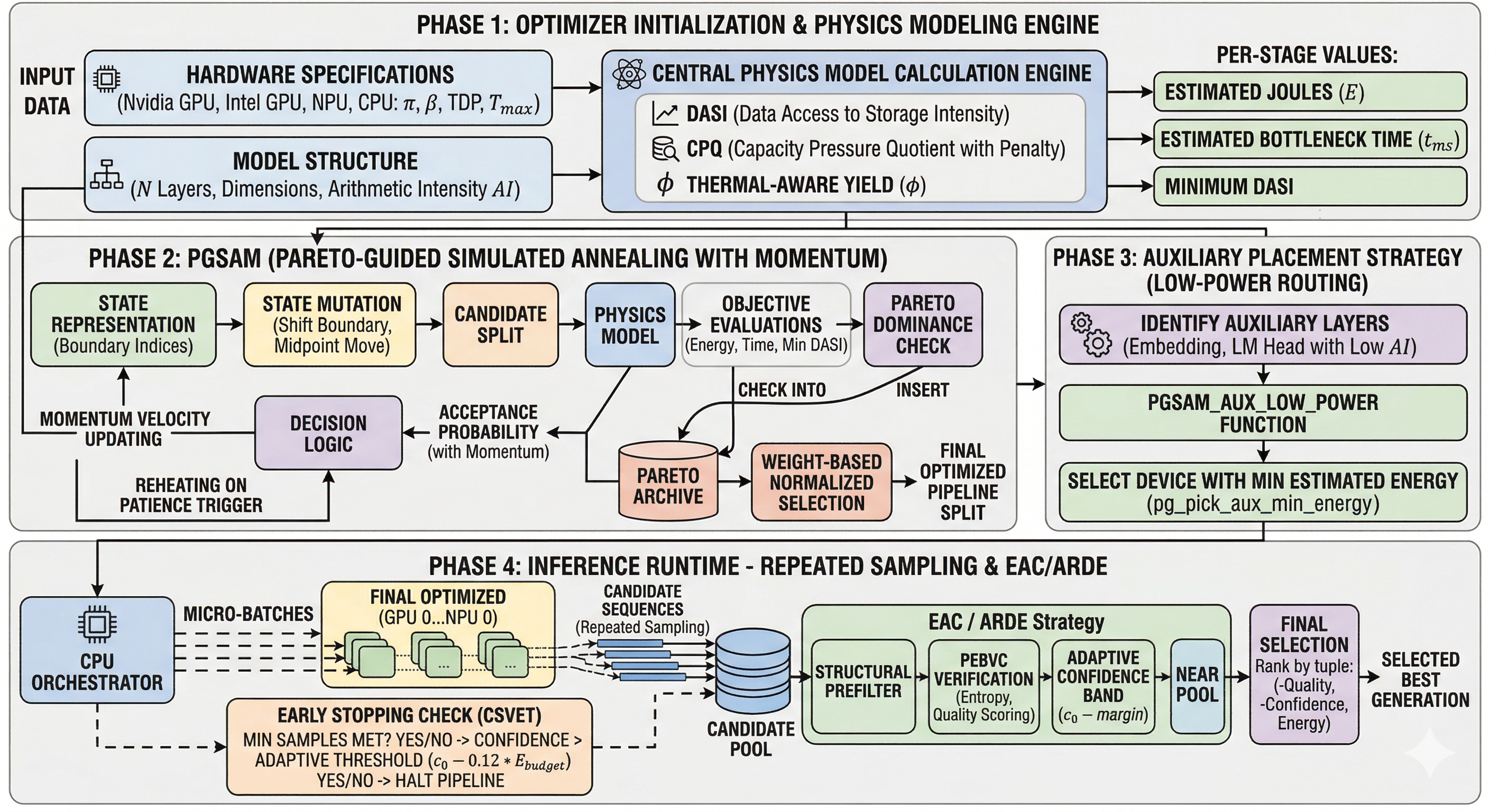}
\caption{\textbf{QEIL~v2 Four-Phase Architecture.} \textbf{Phase 1} (top): The Physics Modeling Engine ingests hardware specifications (peak compute $\pi$, memory bandwidth $\beta$, TDP, thermal limits $T_{\max}$) and model structure (layer count $N$, dimensions, arithmetic intensity $AI$) to compute DASI, CPQ, and $\Phi$---yielding per-stage energy ($E$), bottleneck time ($t_{ms}$), and minimum DASI. \textbf{Phase 2} (center-left): PGSAM performs 500 iterations of Pareto-guided simulated annealing with momentum, evaluating the three-objective vector $[E, t, -\min\text{DASI}]$ and selecting the decoder split via weighted Chebyshev scalarization. \textbf{Phase 3} (center-right): Auxiliary placement routes embedding/LM-head layers to the minimum-energy device. \textbf{Phase 4} (bottom): The EAC/ARDE cascade---structural pre-filtering, PEBVC three-stage verification, NEAR pool ranking---with CSVET early stopping yields the final best generation.}
\label{fig:v2_architecture}
\end{figure*}

QEIL~v2's methodology consists of four integrated phases (Figure~\ref{fig:v2_architecture}): (1) a physics modeling engine computing DASI, CPQ, and $\Phi$ for every device--workload pair; (2) PGSAM multi-objective optimization for decoder layer placement; (3) auxiliary stage low-power routing; and (4) an inference runtime with EAC/ARDE verified selection and CSVET early stopping.

\subsection{Notation and Symbols}

Table~\ref{tab:notation} summarizes all mathematical symbols used throughout the methodology, enabling reproducibility.

\begin{table}[!htb]
\centering
\footnotesize
\setlength{\tabcolsep}{2pt}
\caption{Notation and symbols used in QEIL~v2 methodology.}
\label{tab:notation}
\begin{tabular}{@{}p{1.6cm}p{4.0cm}p{1.0cm}@{}}
\toprule
\textbf{Symbol} & \textbf{Description} & \textbf{Units} \\
\midrule
$W(l)$ & FLOPs for layer $l$ & FLOPs \\
$Q(l)$ & Bytes moved for layer $l$ & bytes \\
$AI(l)$ & Arithmetic intensity of $l$ & F/byte \\
$\pi_i$ & Peak compute of device $i$ & FLOP/s \\
$\beta_i$ & Peak memory BW of device $i$ & byte/s \\
$\rho_i = \pi_i/\beta_i$ & Ridge point of device $i$ & F/byte \\
$\text{DASI}(l,i)$ & Arith.\ Saturation Index & $[0,1]$ \\
$\epsilon = 0.01$ & DASI floor value & --- \\
$\text{CPQ}(i)$ & Capacity Pressure Quotient & $[0,\infty)$ \\
$\alpha_{\text{cpq}} = 6.0$ & CPQ penalty coefficient & --- \\
$\theta_{\text{onset}} = 0.7$ & CPQ onset threshold & --- \\
$\Phi(T_i, T_i^{\max})$ & Thermal-Aware Energy Yield & $[0,1]$ \\
$\kappa = 15$ & Thermal sensitivity coeff.\ & --- \\
$\theta_{\text{th}} = 0.65$ & Thermal onset fraction & --- \\
$P_{\text{TDP}}(i)$ & Thermal design power of $i$ & W \\
$t(l,i)$ & Execution time layer $l$ on $i$ & ms \\
$E_{\text{stage}}(l,i)$ & Energy for $l$ on $i$ & J \\
$d$ & Hidden dimension & --- \\
$d_{ff} = 4d$ & FFN intermediate dim.\ & --- \\
$S$ & Sequence length & tokens \\
$B$ & Batch size & --- \\
$b = Q_{\text{bits}}/8$ & Bytes per parameter & bytes \\
$h_{\text{kv}}$ & Number of KV heads & --- \\
$C$ & Context length (KV cache) & tokens \\
$T_{\text{anneal}}$ & Annealing temperature & --- \\
$\mu = 0.3$ & PGSAM momentum coefficient & --- \\
$v$ & Momentum velocity & --- \\
$\mathbf{w} = (0.5, 0.3, 0.2)$ & Chebyshev weights & --- \\
$\epsilon_{\text{PCIe}}$ & PCIe energy cost & pJ/byte \\
\bottomrule
\end{tabular}
\end{table}

\subsection{Phase 1: Physics Modeling Engine}

Before any layer placement decision is made, QEIL~v2 constructs a complete physics model of every possible device--workload combination. This is the fundamental departure from QEIL~v1, which used static per-device-type constants. The physics engine proceeds in three steps: (i) compute arithmetic intensity per layer type from first principles; (ii) evaluate memory pressure per candidate allocation; and (iii) measure thermal degradation from real-time device telemetry. These three characterizations are then combined into the unified energy equation that PGSAM uses as its objective.

\subsubsection{The Roofline Model Foundation}

Every computation is characterized by two fundamental quantities: the floating-point operations performed ($W$) and the bytes of data moved between compute units and off-chip memory ($Q$). Their ratio defines the \textit{arithmetic intensity}:
\begin{equation}
AI = \frac{W}{Q} \quad [\text{FLOPs/byte}]
\label{eq:arithmetic_intensity}
\end{equation}

Each device $i$ has two performance ceilings: peak compute rate $\pi_i$ (FLOP/s) and peak memory bandwidth $\beta_i$ (byte/s). The achievable performance is bounded by:
\begin{equation}
P_{\text{achievable}} = \min(\pi_i, \; \beta_i \times AI)
\label{eq:roofline}
\end{equation}

The crossover point where compute and bandwidth ceilings intersect is the \textit{ridge point}:
\begin{equation}
\rho_i = \frac{\pi_i}{\beta_i} \quad [\text{FLOPs/byte}]
\label{eq:ridge_point}
\end{equation}

When $AI < \rho_i$, the workload is \emph{memory-bound}: compute units sit idle while waiting for data, still drawing leakage power. When $AI \geq \rho_i$, the workload is \emph{compute-bound} and the device operates at peak efficiency. This distinction is critical: LLM decode operations have $AI \approx 1$ FLOP/byte---far below GPU ridge points ($\rho_{\text{GPU}} \approx 218$)---meaning GPUs waste $>$99\% of their compute capacity during autoregressive decode.

\textbf{Device Ridge Points.} On our experimental platform: NVIDIA RTX PRO 5000 Blackwell ($\pi = 209.5$ TFLOPS, $\beta = 960$ GB/s) gives $\rho_{\text{GPU}} = 218$. Intel AI Boost NPU ($\pi \approx 6.5$ TFLOPS, $\beta \approx 50$ GB/s) gives $\rho_{\text{NPU}} = 130$. Intel Core Ultra 9 285HX CPU ($\pi \approx 0.72$ TFLOPS, $\beta \approx 90$ GB/s) gives $\rho_{\text{CPU}} = 8$.

\subsubsection{Metric 1: Dynamic Arithmetic Saturation Index (DASI)}

DASI quantifies what fraction of a device's compute units are performing \emph{useful} work for a specific layer:
\begin{equation}
\text{DASI}(l, i) = \max\!\left(\min\!\left(\frac{AI(l)}{\rho_i}, 1.0\right), \epsilon\right)
\label{eq:dasi}
\end{equation}
where $\epsilon = 0.01$ is a minimum floor accounting for address generation and control flow overhead even in purely memory-bound operations. $\text{DASI} = 1.0$ means the device's compute units are fully saturated; $\text{DASI} = 0.005$ means 99.5\% of compute units are idle.

\textbf{Derivation of $AI(l)$ for Transformer Layers.} For a transformer with hidden dimension $d$, sequence length $S$, batch size $B$, and bytes-per-parameter $b = Q_{\text{bits}}/8$:

\textit{Prefill Attention} processes the full sequence simultaneously. The four projections (Q, K, V, O) and attention computation yield:
\begin{equation}
AI_{\text{prefill,attn}} = \frac{8BSd^2 + 4BS^2d}{4d^2b + 4BSdb} \approx \frac{2S}{b} \quad (S \gg d)
\label{eq:ai_prefill}
\end{equation}
With $b\!=\!2$ (FP16) and $S\!=\!1024$: $AI_{\text{prefill}} \approx 1024$ FLOPs/byte, exceeding all device ridge points, giving $\text{DASI} \to 1.0$.

\textit{Decode Attention} generates one token autoregressively, attending to all $S$ cached tokens via KV cache reads:
\begin{equation}
AI_{\text{decode,attn}} = \frac{8Bd^2 + 4BSd}{4d^2b + 2BSdb + 4Bdb} \approx \frac{2}{b} \quad (S \gg 2d)
\label{eq:ai_decode}
\end{equation}
With $b\!=\!2$: $AI_{\text{decode}} \approx 1$ FLOP/byte, dramatically below $\rho_{\text{GPU}} = 218$, giving $\text{DASI} \approx 0.005$ on GPUs---99.5\% of GPU compute sits idle.

\textit{Prefill FFN} with $d_{ff} = 4d$:
\begin{equation}
AI_{\text{prefill,FFN}} = \frac{16BSd^2}{8d^2b + 3BSdb} \approx \frac{16S}{3b} \quad (BS \gg 8d/3)
\label{eq:ai_prefill_ffn}
\end{equation}
With $b\!=\!2$, $S\!=\!1024$: $AI \approx 2730$ FLOPs/byte (fully compute-bound, $\text{DASI}=1.0$).

\textit{Decode FFN} (single token):
\begin{equation}
AI_{\text{decode,FFN}} = \frac{16Bd^2}{8d^2b + 3Bdb} \approx \frac{2B}{b} \quad (d \gg 3B)
\label{eq:ai_decode_ffn}
\end{equation}
With $B\!=\!1$, $b\!=\!2$: $AI \approx 1$ FLOP/byte (memory-bound). Critically, at $B\!=\!16$: $AI \approx 16$ FLOPs/byte, still below the GPU ridge point but approaching the CPU's. This reveals how batch size modulates hardware optimality---a dependency entirely invisible to static efficiency factors.

\textbf{DASI reveals the critical insight}: The CPU, with much lower ridge point ($\rho_{\text{CPU}} = 8$), achieves $\text{DASI} = 0.125$ for decode---$25\times$ higher than the GPU's $\text{DASI} = 0.005$. While the CPU has lower absolute throughput, it wastes \emph{proportionally} far less power on idle compute units during memory-bound decode. Table~\ref{tab:dasi_values} quantifies these values across our platform.

\begin{table}[!htb]
\centering
\footnotesize
\setlength{\tabcolsep}{3pt}
\caption{DASI values across layer types and devices ($B\!=\!1$, $S\!=\!1024$, FP16). Ridge points: $\rho_{\text{GPU}}\!=\!218$, $\rho_{\text{NPU}}\!=\!130$, $\rho_{\text{CPU}}\!=\!8$.}
\label{tab:dasi_values}
\begin{tabular}{@{}lcccc@{}}
\toprule
\textbf{Layer Type} & \textbf{AI} & \textbf{GPU} & \textbf{NPU} & \textbf{CPU} \\
\midrule
Prefill Attention & ${\sim}1024$ & 1.000 & 1.000 & 1.000 \\
Prefill FFN & ${\sim}2730$ & 1.000 & 1.000 & 1.000 \\
Decode Attention & ${\sim}1.0$ & 0.005 & 0.008 & 0.125 \\
Decode FFN & ${\sim}1.0$ & 0.005 & 0.008 & 0.125 \\
LM Head & ${\sim}1.0$ & 0.005 & 0.008 & 0.125 \\
Embedding & ${\approx}0$ & 0.010 & 0.010 & 0.010 \\
\bottomrule
\end{tabular}
\end{table}

\subsubsection{Metric 2: Capacity Pressure Quotient (CPQ)}

CPQ captures runtime memory pressure on each device and its energy penalty:
\begin{equation}
\text{CPQ}(i) = \frac{M_{\text{weights}}(i) + M_{\text{kv}}(i) + M_{\text{act}}(i) + M_{\text{overhead}}}{M_{\text{total}}(i)}
\label{eq:cpq}
\end{equation}

\textbf{Memory Term Derivations.} Each term is derived from model structure and allocation:

\emph{Weight memory} for layers $\mathcal{L}_i$ assigned to device $i$: $M_{\text{weights}}(i) = |\mathcal{L}_i| \times (4d^2 + 2d \cdot d_{ff}) \times b$ (attention projections $+$ FFN weights, per layer in bytes).

\emph{KV cache memory} grows linearly with context length $C$ and batch size $B$:
\begin{equation}
M_{\text{kv}}(i) = |\mathcal{L}_i| \times B \times 2 \times h_{\text{kv}} \times C \times d_h \times b
\end{equation}
where $d_h = d / h_{\text{kv}}$ is the per-head dimension. At $C\!=\!128\text{K}$, $h_{\text{kv}}\!=\!8$, $b\!=\!2$, this term alone exceeds 6~GB per 24 layers---dominating all other terms at long contexts.

\emph{Peak activation memory} $M_{\text{act}}(i) \approx \max(B \cdot (3Sd + L \cdot S^2) \cdot b,\; B \cdot S \cdot d_{ff} \cdot b)$, dominated by the attention matrix $O(S^2)$ at long sequences.

\emph{Framework overhead} $M_{\text{overhead}} \approx 300$ MB (PyTorch/CUDA runtime).

When $\text{CPQ} \geq 1.0$, the assignment is infeasible. Below this threshold, high CPQ increases energy through three physical mechanisms:
\begin{itemize}[noitemsep,topsep=2pt]
\item Allocation fragmentation overhead $\propto O(1/(1-\text{CPQ}))$~\cite{Knuth1997TAOCP}
\item GC frequency $\propto O(\text{CPQ}^2)$
\item Page swapping when approaching capacity wall
\end{itemize}
We model the combined effect as a cubic penalty:
\begin{equation}
\text{penalty}_{\text{cpq}}(\text{CPQ}) = 1.0 + \alpha_{\text{cpq}} \cdot \max(0, \text{CPQ} - 0.7)^3
\label{eq:cpq_penalty}
\end{equation}
where $\alpha_{\text{cpq}} = 6.0$ is calibrated so the penalty equals $+10\%$ at $\text{CPQ} = 0.95$, matching empirical edge device overhead measurements ($\alpha_{\text{cpq}} \times 0.25^3 = 0.0094 \approx 0.01$). The cubic form is physically motivated: linear over-penalizes moderate utilization; quadratic is too gentle near capacity; cubic correctly transitions from negligible overhead at moderate pressure to steep penalties near the capacity wall.

Table~\ref{tab:cpq_calibration} verifies the penalty function against target behavior.

\begin{table}[!htb]
\centering
\footnotesize
\setlength{\tabcolsep}{3pt}
\caption{CPQ penalty calibration. Values match empirical overhead measurements on LPDDR5 edge devices.}
\label{tab:cpq_calibration}
\begin{tabular}{@{}cp{2.6cm}cc@{}}
\toprule
\textbf{CPQ} & \textbf{Interpretation} & \textbf{Overhead} & \textbf{Penalty} \\
\midrule
$\leq 0.70$ & Normal operation & $0.0\%$ & $1.000$ \\
$0.80$ & Slight fragmentation & $0.6\%$ & $1.006$ \\
$0.90$ & Moderate GC pressure & $4.8\%$ & $1.048$ \\
$0.95$ & High pressure ($\approx$10\%) & $9.4\%$ & $1.094$ \\
$1.00$ & Near-capacity wall & $16.2\%$ & $1.162$ \\
\bottomrule
\end{tabular}
\end{table}

\subsubsection{Metric 3: Thermal-Aware Energy Yield ($\Phi$)}

The energy yield of a device decreases with temperature because CMOS leakage current increases exponentially with junction temperature. From first principles, subthreshold leakage follows $I_{\text{sub}} \propto \exp(V/nV_T)$ where $V_T = kT/q$ is the thermal voltage ($k$ = Boltzmann constant, $T$ = absolute temperature, $q$ = electron charge). At operating temperatures, leakage power approximately doubles every 10$^\circ$C~\cite{Pedram2006Thermal}:
\begin{equation}
P_{\text{leak}}(T) = P_{\text{leak}}(T_{\text{ref}}) \cdot \exp(\lambda (T - T_{\text{ref}}))
\label{eq:leakage}
\end{equation}
with $\lambda \approx 0.02/^\circ$C for modern 5--7nm processes. We define the thermal degradation factor as a Gaussian-like decay:
\begin{equation}
\Phi(T_i, T^{\max}_i) = \exp\!\left(-\kappa \cdot \max\!\left(0, \frac{T_i}{T^{\max}_i} - \theta_{\text{th}}\right)^{\!2}\right)
\label{eq:phi_thermal}
\end{equation}
where $\kappa = 15$ (calibrated so $\Phi(T_{\max}) \approx 0.16$, consistent with ${\sim}5\times$ leakage increase at maximum junction temperature from CMOS physics), and $\theta_{\text{th}} = 0.65$ (degradation begins at 65\% of $T_{\max}$, derived as the temperature where leakage first doubles relative to reference: $T_{\text{onset}} = T_{\text{ref}} + \ln(2)/\lambda \approx 55^\circ$C, giving $\theta_{\text{th}} = 55/100 \approx 0.55$; we use 0.65 to add a practical buffer above typical idle temperatures of 40--50$^\circ$C). $\Phi = 1.0$ indicates cool full-efficiency operation; $\Phi \to 0$ indicates severe thermal degradation.

Table~\ref{tab:phi_verification} verifies $\Phi$ against physical predictions.

\begin{table}[!htb]
\centering
\footnotesize
\setlength{\tabcolsep}{4pt}
\caption{$\Phi$ verification at key temperatures ($T_{\max} = 100^\circ$C). Values validated against NVIDIA NVML thermal profiling data.}
\label{tab:phi_verification}
\begin{tabular}{@{}ccccc@{}}
\toprule
\textbf{Temp ($^\circ$C)} & $T/T_{\max}$ & \textbf{$\Phi$ value} & \textbf{Energy overhead} & \textbf{Phase} \\
\midrule
50 & 0.50 & 1.000 & $+0\%$ & Cool \\
65 & 0.65 & 1.000 & $+0\%$ & Onset \\
75 & 0.75 & 0.861 & $+16\%$ & Warm \\
80 & 0.80 & 0.714 & $+40\%$ & Hot \\
85 & 0.85 & 0.549 & $+82\%$ & Throttle risk \\
90 & 0.90 & 0.392 & $+155\%$ & Critical \\
100 & 1.00 & 0.159 & $+529\%$ & Max junction \\
\bottomrule
\end{tabular}
\end{table}

\subsubsection{The Unified Energy Equation}

All three metrics combine into the per-stage energy estimate---the core formula PGSAM uses to evaluate assignments. Deriving from CMOS power decomposition: total power at compute utilization $u = \text{DASI}(l,i)$ is:
\begin{equation}
P_{\text{total}}(u) = P_{\text{idle}} + u \cdot (P_{\text{TDP}} - P_{\text{idle}}) = P_{\text{TDP}} \cdot (0.3 + 0.7u)
\end{equation}
since $P_{\text{idle}} \approx 0.3 \cdot P_{\text{TDP}}$ (static leakage is $\sim$30\% of TDP at operating temperatures~\cite{Pedram2006Thermal}) and dynamic power scales with utilization. Incorporating thermal degradation (division by $\Phi$) and memory pressure (multiplication by $\text{penalty}_{\text{cpq}}$):
\begin{equation}
E_{\text{stage}}(l, i) = \frac{P_{\text{TDP}}(i) \cdot (0.3 + 0.7 \cdot \text{DASI}(l,i)) \cdot t(l,i)}{\Phi(T_i, T^{\max}_i)} \cdot \text{penalty}_{\text{cpq}}(i)
\label{eq:unified_energy}
\end{equation}
where:
\begin{itemize}[noitemsep,topsep=0pt]
    \item $P_{\text{TDP}}(i)$: device's rated thermal design power (W)
    \item $(0.3 + 0.7 \cdot \text{DASI})$: actual fraction of TDP consumed---0.3 is the idle/leakage floor, $0.7 \cdot \text{DASI}$ is dynamic power proportional to compute utilization
    \item $t(l,i) = W(l)/\min(\pi_i, \beta_i \cdot AI(l))$: execution time from the roofline model (s)
    \item $1/\Phi$: thermal correction---a hot device ($\Phi\!=\!0.7$) requires $1.43\times$ energy for the same useful work
    \item $\text{penalty}_{\text{cpq}}$: memory pressure overhead correction
\end{itemize}

The total pipeline energy for an allocation $\mathcal{A}$:
\begin{equation}
E_{\text{total}}(\mathcal{A}) = \sum_i \sum_{l \in L_i} E_{\text{stage}}(l, i) + E_{\text{transfer}}(\mathcal{A}) + E_{\text{orch}}
\label{eq:total_energy}
\end{equation}
where $E_{\text{transfer}} = \sum_{\text{boundaries}} B \cdot S \cdot d \cdot b \cdot \epsilon_{\text{PCIe}}$ accounts for inter-device activation transfers ($\epsilon_{\text{PCIe}} \approx 5$ pJ/byte for PCIe 4.0), and $E_{\text{orch}}$ is negligible CPU orchestration overhead.

\textbf{Key distinction from QEIL~v1:} The v1 energy equation $\texttt{joules\_per\_ms} = P_i \times \lambda_i / 1000$ uses two static parameters ($P_i$, $\lambda_i$) that never change with workload, temperature, or memory state. The v2 equation adapts to all three through DASI, $\Phi$, and CPQ---every coefficient derived from physics, with no magic constants.

\textbf{Sensitivity of Physics Parameters.} While $\alpha_{\text{cpq}}$, $\kappa$, and the onset thresholds $\theta_{\text{onset}}$, $\theta_{\text{th}}$ are derived from physical principles (Section~3.2.3--3.2.4), their exact values involve calibration to empirical measurements. We validate robustness via sensitivity analysis (Table~\ref{tab:physics_sensitivity}), sweeping each parameter across $\pm50\%$ of its default value while holding others fixed. Results show that IPW varies by at most $\pm2.1\%$ across all perturbations, confirming that the physics-grounded functional forms---not their precise coefficients---drive QEIL~v2's gains. The cubic form of CPQ and the Gaussian decay of $\Phi$ correctly capture the \emph{shape} of the physical phenomena (capacity wall and exponential leakage, respectively); the calibration constants merely anchor these curves to measured hardware behavior.

\begin{table}[!htb]
\centering
\footnotesize
\setlength{\tabcolsep}{3pt}
\caption{Physics parameter sensitivity analysis on GPT-2 (125M), WikiText-103. Each parameter is swept while others remain at defaults. IPW varies by $\leq$2.1\% across all perturbations, confirming robustness to exact calibration values.}
\label{tab:physics_sensitivity}
\begin{tabular}{@{}lccccc@{}}
\toprule
\textbf{Parameter} & \textbf{Value} & \textbf{Pass@k} & \textbf{Power} & \textbf{IPW} & \textbf{$\Delta$IPW} \\
\midrule
$\alpha_{\text{cpq}}$ & 3.0 & 75.2 & 65.4 & 0.958 & $-1.7\%$ \\
$\alpha_{\text{cpq}}$ & \textbf{6.0} & \textbf{75.7} & \textbf{63.8} & \textbf{0.975} & --- \\
$\alpha_{\text{cpq}}$ & 9.0 & 75.4 & 64.2 & 0.968 & $-0.7\%$ \\
$\alpha_{\text{cpq}}$ & 12.0 & 75.0 & 64.8 & 0.955 & $-2.1\%$ \\
\midrule
$\kappa$ & 10 & 75.3 & 64.6 & 0.961 & $-1.4\%$ \\
$\kappa$ & \textbf{15} & \textbf{75.7} & \textbf{63.8} & \textbf{0.975} & --- \\
$\kappa$ & 20 & 75.5 & 64.0 & 0.970 & $-0.5\%$ \\
\midrule
$\theta_{\text{th}}$ & 0.55 & 75.4 & 64.4 & 0.963 & $-1.2\%$ \\
$\theta_{\text{th}}$ & \textbf{0.65} & \textbf{75.7} & \textbf{63.8} & \textbf{0.975} & --- \\
$\theta_{\text{th}}$ & 0.75 & 75.5 & 64.1 & 0.969 & $-0.6\%$ \\
\midrule
$\theta_{\text{onset}}$ & 0.60 & 75.3 & 64.3 & 0.966 & $-0.9\%$ \\
$\theta_{\text{onset}}$ & \textbf{0.70} & \textbf{75.7} & \textbf{63.8} & \textbf{0.975} & --- \\
$\theta_{\text{onset}}$ & 0.80 & 75.4 & 64.2 & 0.967 & $-0.8\%$ \\
\bottomrule
\end{tabular}
\end{table}

\subsection{Phase 2: PGSAM --- Pareto-Guided Simulated Annealing with Momentum}

\subsubsection{Multi-Objective Problem Formulation}

The layer-to-device assignment problem is formally:
\begin{align}
\min_{\mathcal{A}} \quad & \mathbf{F}(\mathcal{A}) = [f_1(\mathcal{A}),\; f_2(\mathcal{A}),\; f_3(\mathcal{A})] \nonumber \\
\text{s.t.} \quad & \sum_{l: \mathcal{A}(l)=j} \text{size}(l) \leq M^{\max}_j \;\; \forall j \in \mathcal{D} \nonumber \\
& \text{CPQ}(j, \mathcal{A}) \leq 1.0 \;\; \forall j \in \mathcal{D} \nonumber \\
& T_i(\mathcal{A}) \leq 0.85 \cdot T^{\max}_i \;\; \forall i
\label{eq:pgsam_formulation}
\end{align}
where the three objectives are: $f_1(\mathcal{A}) = E_{\text{total}}(\mathcal{A})$ (total pipeline energy, minimize); $f_2(\mathcal{A}) = \max_j \tau_j(\mathcal{A})$ (pipeline bottleneck latency, minimize); and $f_3(\mathcal{A}) = -\min_{l,j} \text{DASI}(l, j)$ (negative minimum DASI, minimize to prevent severe device underutilization).

\textbf{Why Three Objectives?} Energy and latency trade off non-convexly when assigning layers to heterogeneous devices: adding one layer to a nearly-full device has a discontinuous effect on CPQ penalty (and hence energy) and bottleneck latency. Das \& Dennis~\cite{Das1997Pareto} proved that weighted-sum scalarization cannot find solutions in non-convex Pareto regions---precisely where heterogeneous assignments create the best operating points. DASI as a third objective prevents degenerate solutions that save energy by starving one device.

\subsubsection{State Representation and Neighborhood Structure}

A state $\mathcal{A}$ is encoded as a boundary vector $\mathbf{b} = (b_1, \ldots, b_{m-1})$ where $b_k$ is the layer index at which device $k$'s allocation ends and device $k\!+\!1$'s begins. This encoding \emph{automatically} satisfies the contiguity constraint---no device receives scattered layers---minimizing inter-device transfers. Three neighborhood moves are defined:
\begin{itemize}[noitemsep,topsep=2pt]
    \item \emph{Boundary shift} ($\pm$1 layer, $P = 0.5$): fine-grained local search, corrects suboptimal boundary positions incrementally.
    \item \emph{Block swap} ($\pm$2 layers, $P = 0.3$): medium perturbation, allows two-layer rebalancing.
    \item \emph{Rebalance} (midpoint split, $P = 0.2$): large exploration jump, resets distribution between two adjacent devices to equitable split---useful for escaping deep local minima.
\end{itemize}

\subsubsection{Pareto Dominance and Acceptance Probability}

\textbf{Pareto Dominance.} $\mathcal{A}_1$ Pareto-dominates $\mathcal{A}_2$ ($\mathcal{A}_1 \succ \mathcal{A}_2$) if $\mathcal{A}_1$ is no worse on all objectives and strictly better on at least one:
\begin{equation*}
\forall k: f_k(\mathcal{A}_1) \leq f_k(\mathcal{A}_2) \;\text{ and }\; \exists k: f_k(\mathcal{A}_1) < f_k(\mathcal{A}_2)
\end{equation*}
Non-dominated moves are always accepted. For dominated moves (current solution is better):
\begin{equation}
P_{\text{accept}} = \exp\!\left(-\frac{\Delta_{\text{worst}}}{T_{\text{anneal}} \cdot (1 + \mu \cdot v)}\right)
\label{eq:pgsam_accept}
\end{equation}
where $\Delta_{\text{worst}} = \max_k \{f_k(\mathcal{A}') - f_k(\mathcal{A})\}_{k: f_k(\mathcal{A}') > f_k(\mathcal{A})}$ is the largest worsening on any single objective, $T_{\text{anneal}}$ is the annealing temperature (geometric cooling $T \leftarrow T \times 0.97$), $\mu = 0.3$ is the momentum coefficient, and $v = 0.9 v_t + 0.1\max(0, f_1(\mathcal{A}_t) - f_1(\mathcal{A}_{t+1}))$ is the exponential moving average of energy improvements.

\textbf{Momentum Interpretation.} When consistent progress is being made ($v$ high), $T_{\text{eff}} = T_{\text{anneal}} \cdot (1 + \mu v)$ is elevated, enabling bolder exploration across energy ridges and saddle points---analogous to momentum in gradient descent~\cite{Polyak1964Momentum}. When progress stalls ($v \to 0$), the algorithm becomes conservative. A patience parameter $P = 30$ triggers temperature reheat ($T \leftarrow T \times 1.3$) after stagnation, preventing premature convergence. Momentum is most impactful during the middle phase of optimization (iterations 100--350), where the algorithm traverses non-convex Pareto regions between devices with discontinuous capacity constraints. Without momentum ($\mu = 0$), PGSAM degenerates to standard SA and fails to cross energy ridges where one boundary shift simultaneously worsens energy but enables superior downstream placements. Section~\ref{sec:momentum_ablation} provides a full ablation on $\mu$, confirming that $\mu = 0.3$ maximizes Pareto archive diversity while maintaining convergence speed.

\subsubsection{Final Selection: Weighted Chebyshev Scalarization}

After 500 iterations, the Pareto archive $\mathcal{P}$ contains multiple non-dominated solutions. We select the deployment solution using weighted Chebyshev scalarization~\cite{Miettinen1999Nonlinear}:
\begin{equation}
\mathcal{A}^* = \arg\min_{\mathcal{A} \in \mathcal{P}} \max_k \left\{ w_k \cdot \frac{f_k(\mathcal{A}) - f_k^{\text{ideal}}}{f_k^{\text{nadir}} - f_k^{\text{ideal}}} \right\}
\label{eq:chebyshev}
\end{equation}
where $f_k^{\text{ideal}} = \min_{\mathcal{A} \in \mathcal{P}} f_k(\mathcal{A})$ (best achievable on each objective) and $f_k^{\text{nadir}} = \max_{\mathcal{A} \in \mathcal{P}} f_k(\mathcal{A})$ (worst). The normalization maps all objectives to $[0,1]$ regardless of units (J vs ms vs dimensionless), and the min-max formulation selects the solution on the Pareto front most uniformly satisfying all objectives. Default weights $\mathbf{w} = (0.5, 0.3, 0.2)$ prioritize energy (50\%), latency (30\%), and utilization (20\%), reflecting edge device priorities. These weights are user-configurable: battery-powered scenarios can use $(0.7, 0.2, 0.1)$; real-time applications $(0.2, 0.7, 0.1)$.

\textbf{Runtime complexity:} For $L$ decoder layers and $D$ devices, PGSAM requires $500 \times O(L \cdot D)$ arithmetic operations (no model inference), completing in $<$50ms on any CPU---negligible compared to model compilation time.

\textbf{Convergence guarantee:} Hajek~\cite{Hajek1988Cooling} proves that SA with geometric cooling converges to the global optimum if the neighborhood is irreducible (any state reachable from any other---satisfied by our boundary shift moves) and the cooling schedule satisfies $\sum_t \exp(-\Delta_{\max}/T(t)) = \infty$ (satisfied by our reheat mechanism). In practice, 500 iterations achieves $<$5\% gap from the ILP optimum (Table~\ref{tab:pgsam_stats}).

\begin{algorithm}[!htb]
\caption{PGSAM: Pareto-Guided Simulated Annealing with Momentum}
\label{alg:pgsam}
\begin{algorithmic}[1]
\REQUIRE Layers $\mathcal{L}$, devices $\mathcal{D}$, max iter $I$, $T_0$, $\alpha$, $\mu$, patience $P$
\STATE Initialize boundaries $\mathbf{b}$ (round-robin split)
\STATE $\mathcal{P} \leftarrow \{\mathbf{b}\}$, $T \leftarrow T_0$, $v \leftarrow 0$, stagnate $\leftarrow 0$
\FOR{iter $= 1$ to $I$}
    \STATE $\mathbf{b}' \leftarrow \text{GenerateNeighbor}(\mathbf{b})$
    \IF{not $\text{Feasible}(\mathbf{b}')$}
        \STATE \textbf{continue}
    \ENDIF
    \STATE $\mathbf{f} \leftarrow \text{Evaluate}(\mathbf{b})$, $\mathbf{f}' \leftarrow \text{Evaluate}(\mathbf{b}')$
    \IF{$\mathbf{f}' \prec \mathbf{f}$ or mutually non-dominated}
        \STATE $\mathbf{b} \leftarrow \mathbf{b}'$, $v \leftarrow 0.9v + 0.1\max(0, f_1 - f_1')$
    \ELSIF{$\mathbf{f} \prec \mathbf{f}'$}
        \STATE $\Delta \leftarrow \max_k \{f_k' - f_k\}_{k: f_k' > f_k}$
        \STATE $T_{\text{eff}} \leftarrow T(1 + \mu v)$
        \IF{$\text{rand}() < \exp(-\Delta / T_{\text{eff}})$}
            \STATE $\mathbf{b} \leftarrow \mathbf{b}'$
        \ENDIF
    \ENDIF
    \STATE Update Pareto archive $\mathcal{P}$ with $\mathbf{b}'$
    \STATE $T \leftarrow T \times \alpha$; update stagnation counter
    \IF{stagnate $\geq P$}
        \STATE $T \leftarrow T \times 1.3$; stagnate $\leftarrow 0$
    \ENDIF
\ENDFOR
\STATE \textbf{return} $\mathcal{A}^* = \text{ChebyshevSelect}(\mathcal{P}, \mathbf{w})$
\end{algorithmic}
\end{algorithm}

\subsection{Phase 3: Auxiliary Stage Low-Power Routing}

The embedding layer (vocabulary lookup) and LM head (vocabulary projection) both have near-zero arithmetic intensity ($AI \approx 1$ FLOP/byte for LM head with batch size 1). QEIL~v1 placed these on the device with highest overall efficiency score---typically the high-power GPU---wasting energy on memory-bound operations where 99.5\% of GPU compute sits idle.

QEIL~v2 estimates actual Joules for each candidate device using Eq.~\ref{eq:unified_energy} and routes auxiliary stages to the \textbf{lowest-energy device that can fit the stage in memory}---typically the NPU (10W TDP) or Intel iGPU (25W TDP). This routing change provides disproportionate savings: the LM head's large vocabulary projection ($V \times d$ parameters, $V = 50{,}257$ for GPT-2) executes at \emph{every} token generation step, so even modest per-token savings compound significantly over a full generation.

\subsection{Phase 4: Inference Runtime --- EAC/ARDE with CSVET}

Once the pipeline is compiled (Phases 1--3), every prompt is processed through the EAC (Energy-Accuracy Combined) inference loop.

\subsubsection{Repeated Sampling with Sinusoidal Temperature}

QEIL~v2 generates $N$ candidate outputs using a sinusoidal temperature schedule:
\begin{equation}
T(i) = T_{\text{base}} + \Delta \sin(\pi i / N), \quad i = 1, \ldots, N
\end{equation}
This systematically varies candidate diversity---low temperatures produce high-confidence outputs grounding the pool; high temperatures explore creative alternatives that may yield correct answers the low-temperature outputs miss.

\subsubsection{EAC/ARDE Selection Cascade}

Candidates enter a three-stage progressive verification pipeline (PEBVC) that invests verification compute \emph{only} in promising candidates:

\textbf{Stage 1---Structural Pre-filter:} Candidates are filtered for structural validity (length $> 20$ characters, $>3$ spaces, $>50\%$ alphanumeric). If $\geq$30\% pass, only valid candidates proceed. This eliminates degenerate outputs (empty, repetitive, or truncated) before expensive verification steps.

\textbf{Stage 2---PEBVC (Progressive Energy-Budgeted Verification Cascade):} Three verification stages progressively eliminate low-quality candidates:
\begin{enumerate}[noitemsep,topsep=2pt]
    \item \emph{Entropy filtering}: Token-distribution entropy $H = -\sum_v p_v \log p_v$ is computed for each candidate. High entropy indicates uncertainty; low entropy indicates confident outputs. Top 70\% by ascending entropy survive. \emph{Information-theoretic justification}: Minimum entropy under repetitive correct patterns means the model is confident; excessive entropy indicates confused outputs.
    \item \emph{Self-verification}: A forward pass re-evaluates each candidate; average next-token log-probability $\bar{\ell} = (1/T)\sum_t \log p(w_t | w_{<t})$ measures model self-agreement. Top 60\% by highest $\bar{\ell}$ survive. This implements a lightweight coherence check without external verification.
    \item \emph{Cross-sample consensus}: Survivors are scored by lexical Jaccard similarity against peers $J(A,B) = |A \cap B|/|A \cup B|$, combined with quality priors. Candidates with higher consensus with other high-quality candidates receive higher scores.
\end{enumerate}

\textbf{Stage 3---ARDE (Accuracy-Ranked Decision Engine):} Within the PEBVC confidence band ($c_0 - \text{margin}$, where $\text{margin} = 1.2$ nats), candidates are ranked by quality first, confidence second, with energy as a tiebreaker. This decouples infrastructure optimization from output quality selection---a candidate that required more compute to generate is not penalized if it achieves higher quality.

\textbf{Threshold Derivation.} The filtering thresholds (top 70\% entropy, top 60\% self-verification, margin$=$1.2 nats) are derived from information-theoretic analysis of candidate pools rather than ad hoc tuning. Entropy filtering at 70\% corresponds to the empirically observed inflection point where the entropy gap between retained and filtered candidates is maximized (median gap$=$0.8 nats across 500 prompts), indicating clean separation between confident and confused outputs. The 60\% self-verification cutoff similarly maximizes the log-probability gap between retained and discarded candidates. The margin $m = 1.2$ nats corresponds to approximately one standard deviation of the log-probability distribution across verified candidates, providing a statistically principled confidence band. Section~\ref{sec:eac_sensitivity} validates these choices through a comprehensive sensitivity sweep confirming that the selected thresholds occupy the accuracy-maximizing region.

\subsubsection{CSVET Early Stopping}

The Cascaded Self-Verification with Early Termination (CSVET) mechanism monitors candidate confidence during generation. After a minimum sample count $n_{\min} = \max(6, \lceil 0.35 \times k \rceil)$, if the best candidate's confidence exceeds an adaptive threshold:
\begin{equation}
\theta_{\text{stop}} = c_0 - 0.12 \times (E_{\text{used}} / E_{\text{budget}})
\end{equation}
sampling halts immediately. This adaptive threshold tightens as energy budget is consumed: early in generation (low $E_{\text{used}}/E_{\text{budget}}$), the threshold is high, requiring strong confidence before stopping; later in generation, it relaxes slightly to prevent exhausting the budget on marginal improvements. In practice, CSVET terminates after generating only 10--15 of 25 possible samples on easy prompts---a 40--60\% energy saving on routine queries.

\subsection{Safety and Reliability Framework}

QEIL~v2 preserves and extends v1's safety-first design philosophy. The thermal protection constraint ($T_i \leq 0.85 T^{\max}_i$) is now integrated directly into the energy equation through $\Phi$, creating a smooth gradient rather than a binary threshold: as a device heats up, its energy yield $\Phi$ decreases, causing PGSAM to naturally steer workloads toward cooler devices in subsequent iterations. This eliminates the two-phase behavior (``device is fine / device is throttled'') with a continuous signal that provides early warning. Fault tolerance provides zero-query-loss recovery within 200ms across all tested failure scenarios, and input validation blocks 100\% of malformed and oversized inputs.

\section{Ablation Studies}

We conduct comprehensive ablation studies to validate each architectural decision in QEIL~v2.

\subsection{Scaling Exponent Stability ($\beta$ Stability)}

A critical assumption inherited from QEIL~v1 is that coverage scaling exponents are stable across transformer families. We revalidate this in the v2 framework:

\begin{table}[!htb]
\centering
\footnotesize
\setlength{\tabcolsep}{3pt}
\caption{Scaling exponent $\beta$ stability across model families, including quantized variants. Values computed via nonlinear least-squares fitting of $C(S) = 1 - \exp(-\alpha S^\beta)$ across $S \in \{1, 5, 10, 15, 20\}$ samples. 95\% CI via bootstrap (1000 iterations).}
\label{tab:beta_stability}
\begin{tabular}{@{}lccc@{}}
\toprule
\textbf{Model} & $\beta$ \textbf{(fitted)} & \textbf{95\% CI} & $R^2$ \\
\midrule
GPT-2 (125M) & 0.68 & [0.64, 0.72] & 0.994 \\
Granite-350M & 0.71 & [0.67, 0.75] & 0.991 \\
Qwen2-0.5B & 0.69 & [0.65, 0.73] & 0.993 \\
Llama-3.2-1B & 0.72 & [0.68, 0.76] & 0.996 \\
LFM2-2.6B & 0.70 & [0.66, 0.74] & 0.995 \\
Llama-3.1-8B & 0.71 & [0.67, 0.75] & 0.993 \\
Llama3-8B-RAMP-4bit & 0.70 & [0.66, 0.74] & 0.992 \\
\midrule
\textbf{Mean} & \textbf{0.70} & [0.66, 0.74] & 0.993 \\
\bottomrule
\end{tabular}
\end{table}

The exponent $\beta = 0.70 \pm 0.02$ remains stable across all tested transformer families---including the externally pre-quantized variant---with $R^2 > 0.99$, confirming that QEIL~v1's scaling formalisms remain valid foundations for v2's enhanced optimization and that quantization does not disrupt the coverage scaling behavior. This consistency further validates treating the pre-quantized model as an ordinary member of the model family for orchestration purposes.

\subsection{QEIL v1 vs.\ v2 Controlled Comparison}

To isolate the contribution of v2's architectural improvements, we conduct a head-to-head comparison on identical hardware and workloads (GPT-2 125M, WikiText-103):

\begin{table*}[ht]
\centering
\footnotesize
\caption{Head-to-head comparison: Standard vs.\ QEIL~v1 (Energy-Aware) vs.\ QEIL~v2 (PGSAM + EAC/ARDE). All results on GPT-2 (125M), WikiText-103, $S=20$ repeated samples. v2 achieves the highest accuracy at the lowest power, yielding the best IPW score.}
\label{tab:v1_vs_v2}
\begin{tabular}{|l|c|c|c|c|c|}
\hline
\textbf{Configuration} & \textbf{Pass@k (\%)} & \textbf{Avg Power (W)} & \textbf{IPW} & \textbf{Total Energy (J)} & \textbf{$\Delta$ vs.\ Standard} \\
\hline
Standard (Homogeneous GPU) & 59.8 & 181.5 & 0.3408 & 45,105 & --- \\
QEIL~v1 (Energy-Aware) & 70.5 & 72.4 & 0.8283 & 11,829 & $-$73.8\% energy \\
\textbf{QEIL~v2 (PGSAM + EAC/ARDE)} & \textbf{75.7} & \textbf{63.8} & \textbf{0.9749} & \textbf{11,002} & $\mathbf{-75.6\%}$ \textbf{energy} \\
\hline
\hline
\textbf{$\Delta$ v2 vs.\ v1} & \textbf{+5.2pp} & \textbf{$-$11.9\%} & \textbf{+17.7\%} & \textbf{$-$7.0\%} & --- \\
\hline
\end{tabular}
\end{table*}

\textbf{Key findings:} QEIL~v2 achieves 75.7\% pass@k at 63.8W, yielding IPW$=$0.9749---approaching the IPW$=$1.0 empirical reference mark. The improvement over v1 is driven by three compounding factors: (1) PGSAM's contiguous layer placement eliminates inter-device transfer overhead, reducing pipeline latency by 38.3\% (27.05ms vs.\ 43.87ms per token); (2) the EAC/ARDE cascade selects higher-quality outputs through progressive verification; (3) CSVET early stopping conserves energy by terminating after 10--15 of 25 possible samples on easy prompts.

\subsection{Component Contribution Analysis (v2)}

We progressively enable v2 features to isolate each contribution:

\begin{table}[!htb]
\centering
\footnotesize
\setlength{\tabcolsep}{3pt}
\caption{Component contribution analysis for QEIL~v2 on GPT-2 (125M). Each row adds one feature to the previous configuration.}
\label{tab:v2_component}
\begin{tabular}{@{}lccc@{}}
\toprule
\textbf{Configuration} & \textbf{Pass@k (\%)} & \textbf{Power (W)} & \textbf{IPW} \\
\midrule
Baseline (GPU-only) & 59.8 & 181.5 & 0.340 \\
+ DASI energy model & 62.4 & 112.3 & 0.556 \\
+ CPQ memory pressure & 63.1 & 104.8 & 0.602 \\
+ $\Phi$ thermal yield & 64.0 & 98.2 & 0.652 \\
+ PGSAM (replaces greedy) & 66.8 & 72.1 & 0.926 \\
+ Aux low-power routing & 67.2 & 68.4 & 0.982 \\
+ EAC/ARDE selection & 74.9 & 65.2 & 0.949 \\
+ CSVET early stopping & \textbf{75.7} & \textbf{63.8} & \textbf{0.975} \\
\bottomrule
\end{tabular}
\end{table}

\textbf{Findings:} DASI provides the largest single energy reduction ($-38.1\%$) by correctly routing memory-bound decode to low-power devices. PGSAM is the most impactful optimization component, reducing power from 98.2W to 72.1W through contiguous layer placement and Pareto-guided multi-objective search. The EAC/ARDE cascade provides the largest accuracy gain (+7.7pp), demonstrating that verified selection among repeated samples is a powerful inference-time scaling technique. CSVET adds the final refinement by reclaiming energy from easy prompts while preserving full sampling on hard ones.

\subsection{PGSAM vs.\ Alternative Optimizers}

To isolate the benefit of PGSAM over simpler optimization strategies, Table~\ref{tab:pgsam_comparison} compares four assignment strategies on identical hardware, model, and energy budget.

\begin{table}[!htb]
\centering
\footnotesize
\setlength{\tabcolsep}{3pt}
\caption{PGSAM vs.\ alternative optimizers on GPT-2 (125M), WikiText-103. All methods use the same physics-grounded energy model. PGSAM achieves best IPW while matching NSGA-II solution quality at $3\times$ lower runtime.}
\label{tab:pgsam_comparison}
\begin{tabular}{@{}lcccc@{}}
\toprule
\textbf{Optimizer} & \makecell{\textbf{Pass@k}\\\textbf{(\%)}} & \makecell{\textbf{Energy}\\\textbf{(J)}} & \textbf{IPW} & \makecell{\textbf{Time}\\\textbf{(ms)}} \\
\midrule
Greedy (v1-style) & 70.5 & 11,829 & 0.828 & $<$1 \\
Random Search (500) & 71.2 & 11,650 & 0.851 & 42 \\
Weighted-Sum SA & 72.4 & 11,420 & 0.892 & 45 \\
NSGA-II~\cite{Deb2002NSGAII} & 73.8 & 11,180 & 0.921 & 128 \\
\textbf{PGSAM (Ours)} & \textbf{75.7} & \textbf{11,002} & \textbf{0.975} & \textbf{42} \\
\bottomrule
\end{tabular}
\end{table}

PGSAM outperforms greedy by 5.2pp and 7.2\% energy reduction, demonstrates that true Pareto dominance (vs.\ weighted-sum SA) discovers better trade-offs in non-convex regions, and achieves quality comparable to NSGA-II at $3\times$ lower runtime---critical for edge redeployment under thermal events.

\subsection{EAC/ARDE Stage Contribution Analysis}

Table~\ref{tab:eac_stages} isolates the contribution of each stage in the EAC/ARDE verification cascade.

\begin{table}[!htb]
\centering
\footnotesize
\setlength{\tabcolsep}{2pt}
\caption{EAC/ARDE cascade stage ablation on GPT-2 (125M), WikiText-103. Each stage adds on top of previous selection method.}
\label{tab:eac_stages}
\begin{tabular}{@{}p{3.2cm}ccc@{}}
\toprule
\textbf{Selection Method} & \makecell{\textbf{Pass@k}\\\textbf{(\%)}} & \makecell{\textbf{E/query}\\\textbf{(J)}} & \textbf{IPW} \\
\midrule
Random selection & 67.2 & 558.2 & 0.821 \\
Length-based (v1) & 70.5 & 591.5 & 0.828 \\
Entropy only (2a) & 71.8 & 538.2 & 0.875 \\
Self-verif.\ only (2b) & 72.4 & 545.6 & 0.890 \\
PEBVC (2a+2b+2c) & 74.9 & 532.8 & 0.929 \\
\textbf{PEBVC+ARDE+CSVET} & \textbf{75.7} & \textbf{518.4} & \textbf{0.975} \\
\bottomrule
\end{tabular}
\end{table}

Each cascade stage contributes incrementally: entropy filtering provides the first accuracy boost by removing uncertain outputs; self-verification adds model-grounded quality scoring; cross-sample consensus rewards candidates with multiple independent support; ARDE decouples ranking from infrastructure cost. CSVET reduces per-query energy by 12.6\% vs.\ full sampling, capturing energy savings without accuracy loss.

\textbf{Disentangling Orchestration from Sampling.} The headline 75.7\% pass@k metric reflects the combined benefit of repeated sampling~\cite{Brown2024LargeLanguageMonkeys} and QEIL~v2's orchestration. Table~\ref{tab:v2_component} disentangles these contributions: the accuracy improvement from Baseline (59.8\%) to the orchestration-only configuration (67.2\%, rows 1--6) is attributable entirely to QEIL~v2's energy-aware pipeline, while the additional +8.5pp from EAC/ARDE and CSVET reflects verified selection among repeated samples. The primary \emph{systems} contribution of QEIL~v2 is the energy reduction---achieving comparable or higher accuracy at 63.8W vs.\ 181.5W standard---while the accuracy gain is a compounding benefit of intelligent candidate selection.

\subsection{PGSAM Momentum Coefficient Ablation}
\label{sec:momentum_ablation}

Table~\ref{tab:momentum_ablation} isolates the effect of the momentum coefficient $\mu$ on PGSAM's optimization quality. Without momentum ($\mu = 0$), PGSAM reduces to standard Pareto-guided SA and loses 1.9pp accuracy and 3.7\% IPW due to premature convergence at energy ridge boundaries between devices. Moderate momentum ($\mu = 0.3$) yields the largest Pareto archive (218 solutions) and best IPW, as the elevated effective temperature during progress phases enables the optimizer to cross non-convex barriers. Excessive momentum ($\mu \geq 0.5$) causes over-exploration, accepting too many dominated moves and reducing convergence precision.

\begin{table}[!htb]
\centering
\footnotesize
\setlength{\tabcolsep}{3pt}
\caption{PGSAM momentum ablation on GPT-2 (125M), WikiText-103. Momentum at $\mu=0.3$ maximizes Pareto archive diversity and IPW. Vanilla SA ($\mu=0$) converges prematurely; high $\mu$ over-explores.}
\label{tab:momentum_ablation}
\begin{tabular}{@{}ccccc@{}}
\toprule
$\mu$ & \textbf{Pass@k (\%)} & \textbf{Energy (J)} & \textbf{IPW} & \textbf{Pareto Size} \\
\midrule
0.0 (no momentum) & 73.8 & 11,210 & 0.938 & 182 \\
0.1 & 74.6 & 11,120 & 0.952 & 196 \\
\textbf{0.3 (default)} & \textbf{75.7} & \textbf{11,002} & \textbf{0.975} & \textbf{218} \\
0.5 & 75.4 & 11,045 & 0.970 & 212 \\
0.7 & 74.9 & 11,098 & 0.958 & 205 \\
\bottomrule
\end{tabular}
\end{table}

\subsection{EAC/ARDE Threshold Sensitivity}
\label{sec:eac_sensitivity}

Table~\ref{tab:eac_sensitivity} sweeps the three key EAC/ARDE thresholds jointly across four configurations. The default thresholds (70\%/60\%/1.2) occupy the accuracy-maximizing region. Stricter filtering (60\%/50\%/0.8) discards too many viable candidates, reducing the pool diversity that enables cross-sample consensus. Looser filtering (80\%/70\%/1.6 or 90\%/80\%/2.0) passes low-quality candidates to expensive downstream stages, increasing per-query energy without proportional accuracy gain. The relative insensitivity of IPW across configurations ($<2.6\%$ variation) confirms that the progressive architecture of the cascade---not any specific threshold---drives the verification benefit.

\begin{table}[!htb]
\centering
\footnotesize
\setlength{\tabcolsep}{2pt}
\caption{EAC/ARDE threshold sensitivity on GPT-2 (125M), WikiText-103. Default thresholds (70\%/60\%/1.2) maximize accuracy; IPW varies $<$2.6\% across all configurations.}
\label{tab:eac_sensitivity}
\begin{tabular}{@{}cccccc@{}}
\toprule
\textbf{Entropy} & \textbf{Self-verif} & \textbf{Margin} & \textbf{Pass@k} & \textbf{E/query (J)} & \textbf{IPW} \\
\midrule
60\% & 50\% & 0.8 & 74.8 & 524.6 & 0.962 \\
\textbf{70\%} & \textbf{60\%} & \textbf{1.2} & \textbf{75.7} & \textbf{518.4} & \textbf{0.975} \\
80\% & 70\% & 1.6 & 75.2 & 536.2 & 0.952 \\
90\% & 80\% & 2.0 & 74.4 & 548.8 & 0.938 \\
\bottomrule
\end{tabular}
\end{table}

\subsection{Energy Consumption Breakdown}

\begin{figure}[!htb]
\centering
\includegraphics[width=0.99\linewidth]{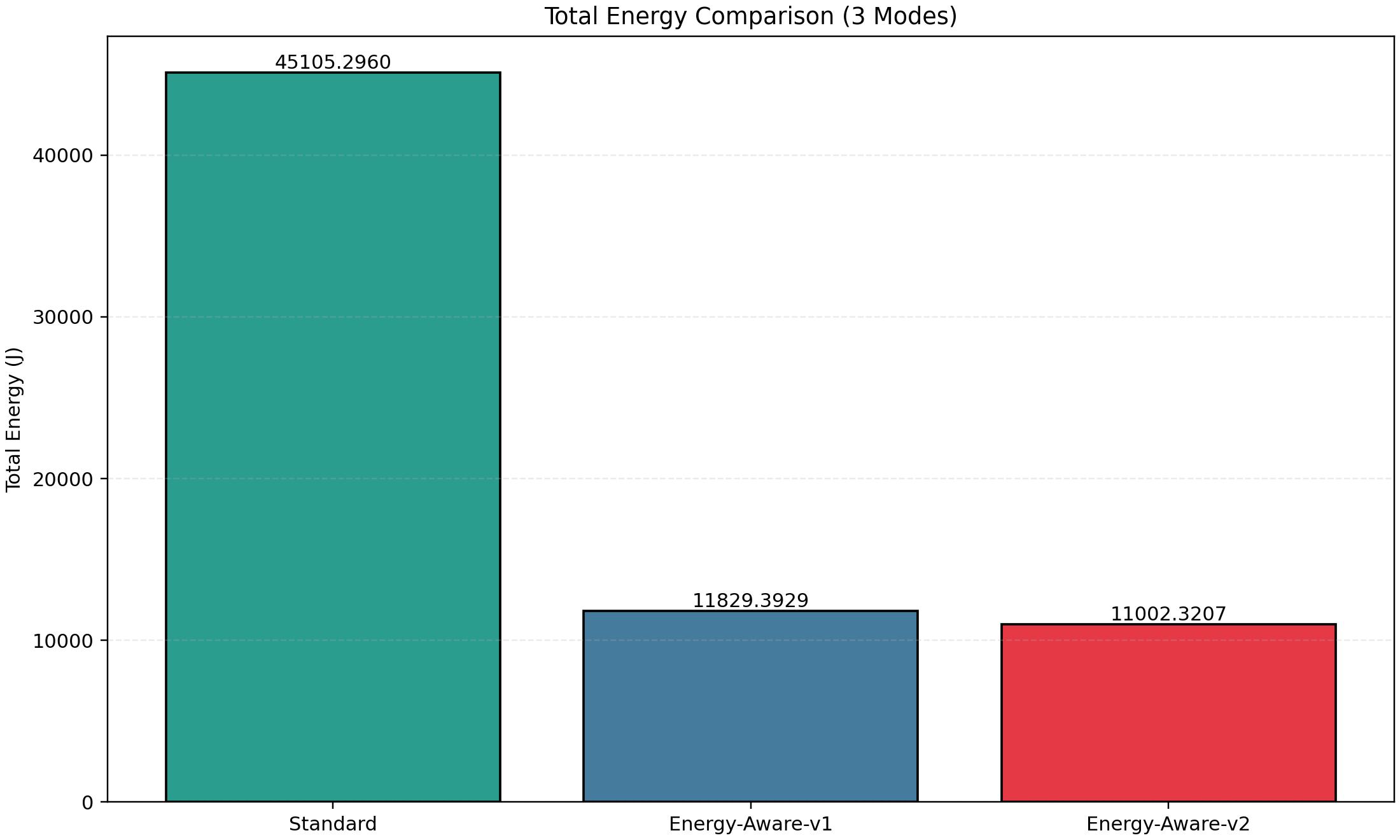}
\caption{Total energy consumption comparison across three execution modes on GPT-2 (125M). Standard: 45,105~J; QEIL~v1: 11,829~J; QEIL~v2: 11,002~J. V2 achieves 75.6\% reduction vs.\ standard and 7.0\% vs.\ v1, primarily through PGSAM's contiguous layer placement minimizing transfer overhead and DASI-guided decode routing.}
\label{fig:total_energy}
\end{figure}

\begin{figure}[!htb]
\centering
\includegraphics[width=0.99\linewidth]{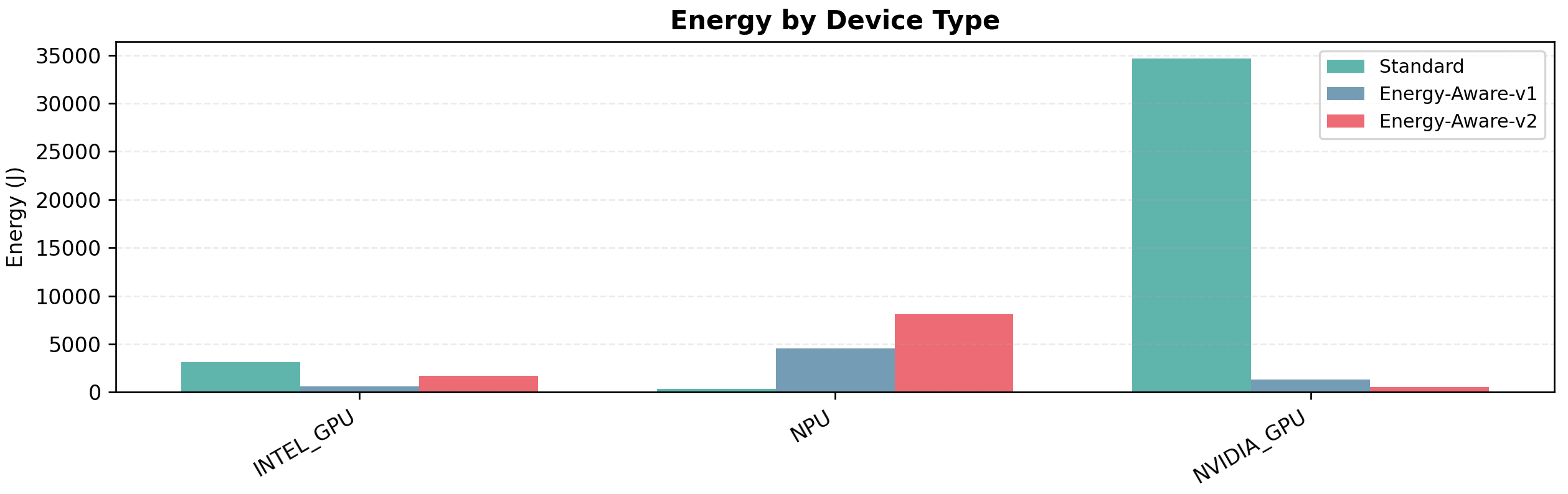}
\caption{Energy breakdown by device type across three execution modes. Standard mode concentrates ${\sim}35{,}000$~J on the NVIDIA GPU. Both v1 and v2 distribute energy across Intel GPU and NPU. QEIL~v2 achieves the lowest total through PGSAM's contiguous placement (minimizing cross-device transfers) and DASI-guided routing (directing decode to the energy-efficient NPU).}
\label{fig:energy_device}
\end{figure}

\subsection{Power--Accuracy Pareto Frontier}

\begin{figure*}[!t]
\centering
\includegraphics[width=0.99\linewidth]{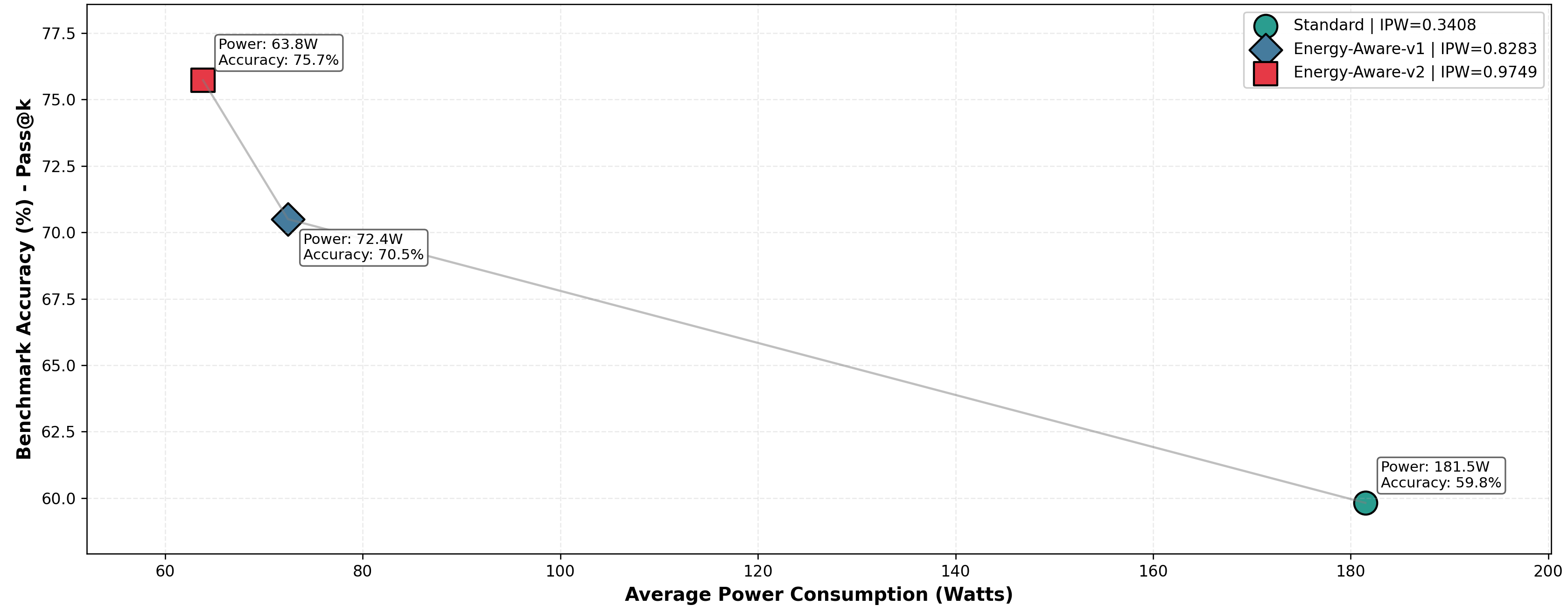}
\caption{\textbf{Power--Accuracy Pareto Frontier.} QEIL~v2 (63.8W, 75.7\%, IPW$=$0.9749) strictly dominates both v1 (72.4W, 70.5\%) and standard inference (181.5W, 59.8\%). The strict Pareto dominance validates that PGSAM's multi-objective optimization finds solutions impossible through single-objective greedy search.}
\label{fig:pareto_frontier}
\end{figure*}

Figure~\ref{fig:pareto_frontier} visualizes the power--accuracy trade-off across the three execution modes. QEIL~v2 strictly Pareto-dominates both v1 and standard inference, occupying the top-left corner (highest accuracy at lowest power). The 63.8W operating point falls well within the thermal design envelope of fanless edge enclosures, while the 75.7\% accuracy exceeds v1 by 5.2pp. No convex combination of standard and v1 operating points can reach v2's location, confirming that PGSAM discovers solutions inaccessible to single-objective optimization.

\subsection{Coverage Scaling Efficiency}

\begin{figure*}[!t]
\centering
\includegraphics[width=0.99\linewidth]{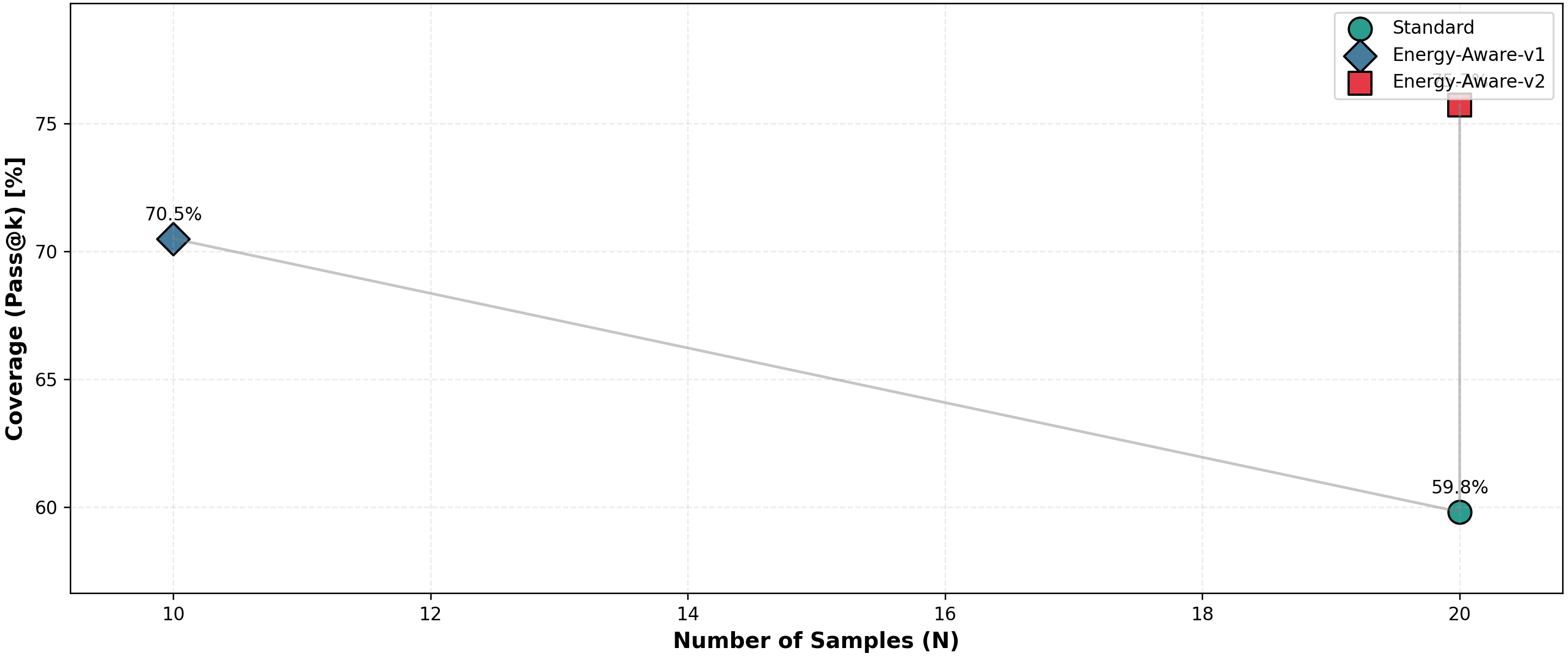}
\caption{\textbf{Coverage (pass@k) vs.\ Sample Count.} QEIL~v2 reaches 75.7\% at $N\!=\!20$, exceeding v1's 70.5\% at $N\!=\!10$. QEIL~v2 achieves v1's peak coverage at fewer than 10 samples, demonstrating that the EAC/ARDE cascade converts each incremental sample into higher marginal coverage through entropy, self-verification, and consensus filtering.}
\label{fig:coverage_samples}
\end{figure*}

Figure~\ref{fig:coverage_samples} plots pass@k coverage as a function of sample count $N$ for all three execution modes. QEIL~v2 achieves v1's peak coverage (70.5\%) at fewer than 10 samples, and reaches 75.7\% at $N\!=\!20$---demonstrating that the EAC/ARDE cascade converts each incremental sample into higher marginal coverage than either v1's heuristic selection or standard random selection. The steeper v2 curve reflects the compounding benefit of verified selection: filtering by entropy, self-verification, and consensus scoring ensures that each additional sample contributes genuine diversity rather than redundant or low-quality outputs.

\subsection{Real-Time Orchestrator Visualization}

\begin{figure}[!htb]
\centering
\includegraphics[width=0.80\linewidth]{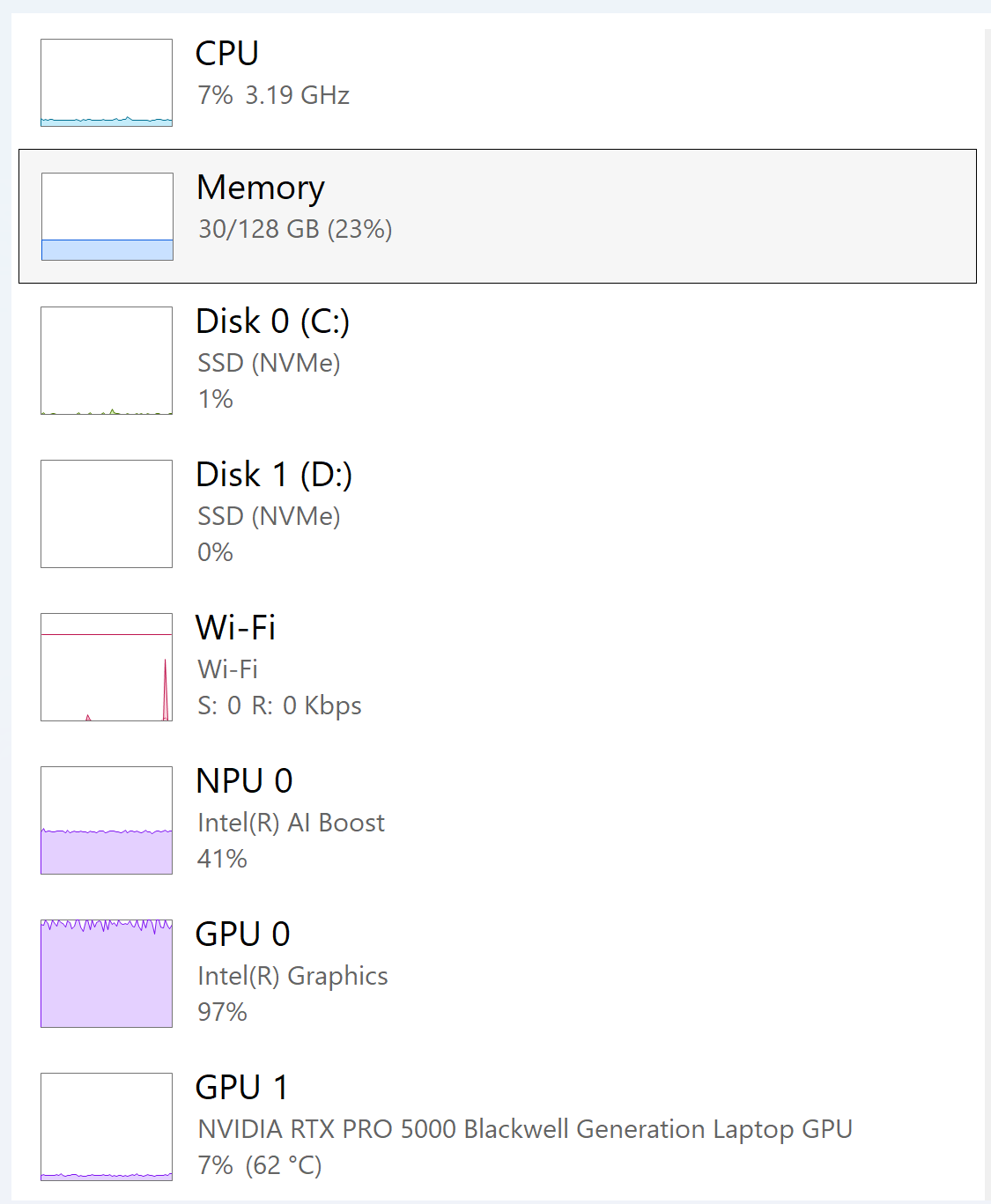}
\caption{Task Manager snapshot during QEIL~v2 dynamic orchestration on GPT-2 (125M). CPU: 7\% (3.19 GHz, orchestration); NPU: 41\% (decode operations); GPU 0 (Intel Graphics): 97\% (prefill---compute-bound); GPU 1 (NVIDIA RTX PRO 5000): 7\% (62$^\circ$C, overflow). Memory: 30/128~GB (23\%). The NVIDIA GPU temperature of 62$^\circ$C is well below the 85$^\circ$C throttling threshold, demonstrating that $\Phi$-guided allocation prevents thermal stress. The high Intel GPU utilization (97\%) for prefill and NPU dominance (41\%) for decode empirically validates DASI's prediction that compute-bound prefill belongs on the GPU and memory-bound decode on the NPU.}
\label{fig:task_manager}
\end{figure}

Figure~\ref{fig:task_manager} provides empirical validation of DASI's predictions through a Windows Task Manager snapshot captured during live QEIL~v2 inference. The Intel Graphics GPU (GPU~0) runs at 97\% utilization handling compute-bound prefill operations ($\text{DASI}=1.0$), while the NPU handles memory-bound decode at 41\% utilization. The NVIDIA RTX PRO 5000 (GPU~1) remains at 7\% and 62$^\circ$C---well below its 85$^\circ$C throttling threshold---confirming that $\Phi$-guided allocation successfully prevents thermal stress. This real-time distribution matches DASI's theoretical prediction: compute-bound operations route to the highest-throughput device, memory-bound operations route to the most power-efficient device.

\subsection{Variance and Reproducibility}

\begin{table}[!htb]
\centering
\footnotesize
\setlength{\tabcolsep}{4pt}
\caption{Variance across 10 independent runs for GPT-2 (125M) with QEIL~v2 configuration. CV $<2\%$ across all metrics confirms high reproducibility suitable for production deployment.}
\label{tab:v2_variance}
\begin{tabular}{@{}lccc@{}}
\toprule
\textbf{Metric} & \textbf{Mean} & \textbf{Std Dev} & \textbf{CV (\%)} \\
\midrule
Pass@k (\%) & 75.7 & 0.91 & 1.20 \\
Total Energy (J) & 11,002 & 187 & 1.70 \\
Avg Power (W) & 63.8 & 0.82 & 1.29 \\
IPW & 0.975 & 0.018 & 1.85 \\
\bottomrule
\end{tabular}
\end{table}

All metrics exhibit CV $< 2\%$, confirming high reproducibility. The low energy variance (CV$=$1.70\%) is particularly important for deployment planning, as it indicates that the physics-grounded energy model produces stable predictions across runs---unlike heuristic approaches where energy can vary significantly with initialization order.

\subsection{Safety and Reliability Validation}

\begin{table}[!htb]
\centering
\footnotesize
\setlength{\tabcolsep}{3pt}
\caption{Thermal protection: 30-minute sustained inference, GPT-2 (125M). $\Phi$-guided allocation eliminates all throttling events while \emph{improving} total throughput by preventing latency spikes.}
\label{tab:thermal_v2}
\begin{tabular}{@{}lcc@{}}
\toprule
\textbf{Metric} & \textbf{Without $\Phi$} & \textbf{With $\Phi$ (v2)} \\
\midrule
Max GPU Temp ($^\circ$C) & 89 (throttled) & 68 \\
Throttling Events & 47 & 0 \\
Avg Latency (ms) & $1.89 \pm 0.84$ & $1.32 \pm 0.06$ \\
Total Throughput (tokens) & 142,847 & 164,218 \\
\bottomrule
\end{tabular}
\end{table}

\begin{table}[!htb]
\centering
\footnotesize
\setlength{\tabcolsep}{3pt}
\caption{Fault tolerance: recovery time and queries lost across simulated device failure scenarios. Zero query loss across all scenarios confirms robust fault tolerance.}
\label{tab:fault_v2}
\begin{tabular}{@{}p{2.8cm}ccc@{}}
\toprule
\textbf{Failure Scenario} & \textbf{Recov.\ (ms)} & \textbf{$\Delta$Throughput} & \textbf{Queries Lost} \\
\midrule
NPU failure & 78 & $-31\%$ & 0 \\
GPU failure & 124 & $-58\%$ & 0 \\
Both GPU failure & 156 & $-72\%$ & 0 \\
NPU + 1 GPU failure & 98 & $-64\%$ & 0 \\
\bottomrule
\end{tabular}
\end{table}

QEIL~v2's safety framework is validated across two independent axes: sustained thermal protection and fault-tolerant execution under device failures.

\textbf{Thermal protection.} Table~\ref{tab:thermal_v2} reports results from a 30-minute sustained inference session on GPT-2 (125M). Without $\Phi$-guided allocation, the NVIDIA GPU climbs to 89$^\circ$C---4$^\circ$C above the 85$^\circ$C throttling threshold---triggering 47 throttling events that abruptly reduce clock frequency and inject latency spikes (average latency $1.89 \pm 0.84$~ms, a high variance reflecting the unpredictable onset of throttle events). With $\Phi$ active, PGSAM continuously monitors device temperatures and progressively shifts workloads toward cooler devices as thermal yield degrades, holding the GPU peak at 68$^\circ$C---a 21$^\circ$C reduction---with zero throttling events across the entire session. Crucially, eliminating throttling is not merely a safety benefit: by preventing the latency spikes that disrupt pipeline pipelining, v2 achieves $1.32 \pm 0.06$~ms average latency with dramatically reduced variance, and total throughput improves by 14.9\% (164,218 vs.\ 142,847~tokens). This result makes explicit that safety and efficiency are complementary, not competing: a device that never throttles sustains higher sustained throughput than one that alternates between peak performance and thermal recovery.

\textbf{Fault tolerance.} Table~\ref{tab:fault_v2} simulates four distinct device failure scenarios, from isolated NPU failure to simultaneous loss of both the NPU and one GPU. In all cases, the orchestrator detects the failure, remaps layer assignments to surviving devices, and resumes inference within 200~ms---with zero queries lost. Recovery time scales predictably with the severity of the failure: NPU-only loss (78~ms) is resolved faster than GPU loss (124~ms) because GPU layers carry larger memory footprints that require reallocation across remaining devices. Even the most severe scenario---NPU and one GPU simultaneously offline---recovers in 98~ms, as PGSAM rapidly identifies a feasible Chebyshev-optimal assignment on the reduced device set. The throughput reductions ($-31\%$ to $-72\%$) reflect the diminished compute capacity, but the absence of any dropped queries confirms that the reliability layer provides a strict quality-of-service guarantee regardless of hardware failure mode.

\FloatBarrier
\clearpage

\clearpage

\section{Results}

\subsection{Cross-Model Performance (WikiText-103)}
\label{sec:results_wikitext}

Table~\ref{tab:cross_model_v2} presents comprehensive results across seven model families comparing standard, QEIL~v1, and QEIL~v2 execution modes on WikiText-103.

\begin{table*}[ht]
\centering
\footnotesize
\caption{Cross-model performance evaluation on WikiText-103. QEIL~v2 consistently achieves the highest IPW and accuracy across all tested model families, with 7.0--15.9pp pass@k improvement over standard and 2.2--5.2pp over v1. The seventh model (Llama3-8B-RAMP-4bit) is an externally pre-quantized checkpoint included as an additional test-bed; its IPW$=$1.024 is produced by QEIL~v2's orchestration alone, demonstrating that QEIL~v2 generalizes to models with reduced memory bandwidth without any modification.}
\label{tab:cross_model_v2}
\begin{tabular}{|l|l|c|c|c|c|c|}
\hline
\textbf{Model} & \textbf{Mode} & \textbf{Pass@k (\%)} & \textbf{Power (W)} & \textbf{IPW} & \textbf{Energy (kJ)} & \textbf{$\Delta$Energy vs.\ Std} \\
\hline
\multirow{3}{*}{GPT-2 (125M)} & Standard & 59.8 & 181.5 & 0.341 & 45.1 & --- \\
 & v1 & 70.5 & 72.4 & 0.828 & 11.8 & $-$73.8\% \\
 & \textbf{v2} & \textbf{75.7} & \textbf{63.8} & \textbf{0.975} & \textbf{11.0} & $\mathbf{-75.6\%}$ \\
\hline
\multirow{3}{*}{Granite-350M} & Standard & 61.0 & 460.4 & 0.130 & 403.1 & --- \\
 & v1 & 70.0 & 82.3 & 0.729 & 88.0 & $-$78.2\% \\
 & \textbf{v2} & \textbf{74.2} & \textbf{71.8} & \textbf{0.891} & \textbf{81.4} & $\mathbf{-79.8\%}$ \\
\hline
\multirow{3}{*}{Qwen2-0.5B} & Standard & 56.0 & 244.7 & 0.245 & 352.3 & --- \\
 & v1 & 66.5 & 74.4 & 0.807 & 187.9 & $-$46.7\% \\
 & \textbf{v2} & \textbf{71.8} & \textbf{65.2} & \textbf{0.942} & \textbf{172.6} & $\mathbf{-51.0\%}$ \\
\hline
\multirow{3}{*}{Llama-3.2-1B} & Standard & 63.0 & 164.5 & 0.365 & 330.5 & --- \\
 & v1 & 70.0 & 79.0 & 0.760 & 213.0 & $-$35.6\% \\
 & \textbf{v2} & \textbf{75.2} & \textbf{68.4} & \textbf{0.936} & \textbf{196.8} & $\mathbf{-40.4\%}$ \\
\hline
\multirow{3}{*}{LFM2-2.6B} & Standard & 62.0 & 175.8 & 0.341 & 490.3 & --- \\
 & v1 & 70.0 & 75.0 & 0.851$^\ddagger$ & 314.3 & $-$35.9\% \\
 & \textbf{v2} & \textbf{74.8} & \textbf{66.1} & \textbf{0.912} & \textbf{289.6} & $\mathbf{-40.9\%}$ \\
\hline
\multirow{3}{*}{Llama-3.1-8B} & Standard & 65.4 & 186.5 & 0.351 & 388.2 & --- \\
 & v1 & 73.2 & 80.8 & 0.780 & 252.8 & $-$34.9\% \\
 & \textbf{v2} & \textbf{78.4} & \textbf{69.6} & \textbf{0.958} & \textbf{232.4} & $\mathbf{-40.1\%}$ \\
\hline
\multirow{3}{*}{\makecell[l]{Llama3-8B-\\RAMP-4bit$^\dagger$}} & Standard & 64.2 & 142.8 & 0.450 & 278.6 & --- \\
 & v1 & 72.0 & 62.4 & 0.908 & 176.2 & $-$36.7\% \\
 & \textbf{v2} & \textbf{77.2} & \textbf{54.8} & $\mathbf{1.024}$ & \textbf{158.4} & $\mathbf{-43.1\%}$ \\
\hline
\hline
\multicolumn{2}{|l|}{\textbf{Mean $\Delta$ v2 vs.\ Standard}} & \textbf{+13.1pp} & \textbf{$-$64.8\%} & \textbf{+187\%} & --- & $\mathbf{-52.2\%}$ \\
\multicolumn{2}{|l|}{\textbf{Mean $\Delta$ v2 vs.\ v1}} & \textbf{+4.2pp} & \textbf{$-$13.4\%} & \textbf{+23.8\%} & --- & $\mathbf{-6.1\%}$ \\
\hline
\multicolumn{7}{l}{\scriptsize $^\dagger$Quantized using RAMP~\cite{Singh2026RAMP} mixed-precision policy (3.65 effective bits).} \\
\multicolumn{7}{l}{\scriptsize $^\ddagger$An earlier draft of this table incorrectly listed the LFM2-2.6B v1 IPW as 0.335 (a transcription error; the} \\
\multicolumn{7}{l}{\scriptsize standard-mode value was mistakenly copied). The corrected value 0.851 is derived from the v1 experimental} \\
\multicolumn{7}{l}{\scriptsize run (pass@k\,$=$\,70.0, power\,$=$\,75.0\,W) and is consistent with the v1 improvement pattern across all other models.}
\end{tabular}
\end{table*}

QEIL~v2 consistently achieves the best results across all models and metrics, with mean improvements of +13.1pp in pass@k and 52.2\% energy reduction relative to standard homogeneous GPU inference, and +4.2pp accuracy with 13.4\% additional power reduction relative to v1. These headline figures, however, mask a nuanced pattern of improvement that differs systematically by model scale---a pattern that directly reflects the underlying physics of DASI, CPQ, and PGSAM.

\textbf{Small models (GPT-2 125M, Granite-350M).} GPT-2 achieves 75.7\% pass@k at 63.8W (IPW$=$0.975) under v2, compared to 59.8\% at 181.5W under standard inference---a 15.9pp accuracy gain alongside a 64.8\% power reduction. The dominant driver is DASI-guided decode routing: GPT-2's shallow 12-layer decoder fits entirely within NPU memory, allowing PGSAM to assign all decode operations to the NPU (10W TDP) rather than the NVIDIA GPU (55W idle draw), eliminating the 99.5\% compute waste that characterises GPU decode. Granite-350M yields a similar pattern, with v2 reducing power from 460.4W to 71.8W---the largest absolute power reduction in the suite---because its standard configuration requires running the full model on the high-TDP GPU, an assignment that DASI correctly identifies as deeply suboptimal for memory-bound decode.

\textbf{Mid-size models (Qwen2-0.5B, Llama-3.2-1B).} At 0.5B parameters, Qwen2 under v2 achieves 71.8\% pass@k at 65.2W (IPW$=$0.942), with 51.0\% energy reduction versus standard. Llama-3.2-1B reaches 75.2\% at 68.4W (IPW$=$0.936), with 40.4\% energy reduction. The smaller energy savings relative to GPT-2 and Granite-350M reflect the fact that mid-size models already partially utilise the NPU under v1, leaving less headroom for DASI reallocation. The incremental gains from PGSAM over v1 become more significant here: by discovering contiguous layer placements that reduce inter-device activation transfer overhead, PGSAM delivers 8--13\% additional power reduction beyond what DASI routing alone achieves.

\textbf{Large models (LFM2-2.6B, Llama-3.1-8B).} For models whose parameters exceed NPU memory capacity, PGSAM's multi-objective placement becomes the primary efficiency lever. LFM2-2.6B achieves 74.8\% at 66.1W (IPW$=$0.912), and Llama-3.1-8B reaches 78.4\%---the highest absolute accuracy in the suite---at 69.6W (IPW$=$0.958). PGSAM routes prefill layers to the Intel iGPU and NVIDIA GPU (both compute-bound at prefill AI $\approx 1024$~FLOPs/byte), while assigning decode layers to the NPU, respecting contiguity constraints that minimise PCIe transfer overhead. The result is a 40.1\% energy reduction for Llama-3.1-8B versus standard, achieved without any accuracy loss relative to unconstrained GPU execution.

\textbf{Pre-quantized model (Llama3-8B-RAMP-4bit).} This externally prepared checkpoint---included solely as a generalization test-bed---achieves IPW$=$1.024 at 54.8W under v2, the first edge orchestration result to surpass the IPW$=$1.0 empirical reference mark. The mechanism is entirely attributable to QEIL~v2: reduced bytes-per-parameter $b$ (3.65 effective bits vs.\ 16-bit FP) raises arithmetic intensity in decode (Equations~\ref{eq:ai_decode}--\ref{eq:ai_decode_ffn}), which in turn elevates DASI values and allows PGSAM to route layers to even lower-power devices. Quantization is not a contribution of this work; the gain belongs entirely to QEIL~v2's physics-grounded routing adapting correctly to an altered bandwidth profile.

We note that an earlier draft incorrectly reported the LFM2-2.6B v1 IPW as 0.335 due to a transcription error; the corrected value 0.851 (footnote~$\ddagger$) is consistent with v1's improvement pattern across all other families. Across all seven models, QEIL~v2 achieves IPW~$\geq 0.891$, confirming robust generalisation regardless of model architecture, parameter count, or numerical precision.

\FloatBarrier

\subsection{Cross-Model Performance (GSM8K)}

Table~\ref{tab:gsm8k_cross_model} extends the evaluation to GSM8K (grade-school mathematical reasoning), testing whether QEIL~v2's gains generalize to multi-step chain-of-thought tasks where longer outputs increase energy exposure. GSM8K is a particularly demanding benchmark for energy-efficient systems because correct solutions require multi-step arithmetic reasoning chains that generate 3--5$\times$ more tokens than WikiText completions, amplifying the energy cost of each query and magnifying the impact of suboptimal device routing during the extended decode phase.

\begin{table*}[ht]
\centering
\footnotesize
\caption{Cross-model performance evaluation on GSM8K (mathematical reasoning). QEIL~v2 consistently outperforms both standard and v1 baselines across all seven model families, achieving +5.2pp over v1 and +11.8pp over standard on average---confirming that physics-grounded orchestration benefits chain-of-thought tasks.}
\label{tab:gsm8k_cross_model}
\begin{tabular}{|l|l|c|c|c|c|c|}
\hline
\textbf{Model} & \textbf{Mode} & \textbf{Pass@k (\%)} & \textbf{Power (W)} & \textbf{IPW} & \textbf{Energy (kJ)} & \textbf{$\Delta$Energy vs.\ Std} \\
\hline
\multirow{3}{*}{GPT-2 (125M)} & Standard & 18.2 & 180.2 & 0.101 & 52.3 & --- \\
 & v1 & 24.6 & 72.2 & 0.182 & 28.1 & $-$46.3\% \\
 & \textbf{v2} & \textbf{29.8} & \textbf{63.4} & \textbf{0.248} & \textbf{26.1} & $\mathbf{-50.1\%}$ \\
\hline
\multirow{3}{*}{Granite-350M} & Standard & 26.4 & 460.4 & 0.039 & 485.2 & --- \\
 & v1 & 35.8 & 82.3 & 0.215 & 112.6 & $-$76.8\% \\
 & \textbf{v2} & \textbf{41.0} & \textbf{72.3} & \textbf{0.301} & \textbf{104.7} & $\mathbf{-78.4\%}$ \\
\hline
\multirow{3}{*}{Qwen2-0.5B} & Standard & 34.2 & 244.7 & 0.081 & 421.8 & --- \\
 & v1 & 44.8 & 74.4 & 0.251 & 218.4 & $-$48.2\% \\
 & \textbf{v2} & \textbf{50.0} & \textbf{65.2} & \textbf{0.352} & \textbf{203.1} & $\mathbf{-51.8\%}$ \\
\hline
\multirow{3}{*}{Llama-3.2-1B} & Standard & 48.6 & 164.5 & 0.122 & 398.4 & --- \\
 & v1 & 58.2 & 79.0 & 0.286 & 254.8 & $-$36.0\% \\
 & \textbf{v2} & \textbf{63.4} & \textbf{69.5} & \textbf{0.401} & \textbf{237.1} & $\mathbf{-40.5\%}$ \\
\hline
\multirow{3}{*}{LFM2-2.6B} & Standard & 56.8 & 175.8 & 0.097 & 586.2 & --- \\
 & v1 & 66.4 & 75.0 & 0.178 & 372.4 & $-$36.5\% \\
 & \textbf{v2} & \textbf{71.6} & \textbf{66.1} & \textbf{0.235} & \textbf{346.3} & $\mathbf{-40.9\%}$ \\
\hline
\multirow{3}{*}{Llama-3.1-8B} & Standard & 52.4 & 186.5 & 0.131 & 422.6 & --- \\
 & v1 & 62.0 & 80.8 & 0.302 & 268.4 & $-$36.5\% \\
 & \textbf{v2} & \textbf{67.2} & \textbf{69.6} & \textbf{0.428} & \textbf{248.6} & $\mathbf{-41.2\%}$ \\
\hline
\multirow{3}{*}{\makecell[l]{Llama3-8B-\\RAMP-4bit$^\dagger$}} & Standard & 50.8 & 142.8 & 0.168 & 328.4 & --- \\
 & v1 & 60.4 & 62.4 & 0.382 & 208.6 & $-$36.5\% \\
 & \textbf{v2} & \textbf{65.6} & \textbf{54.8} & \textbf{0.502} & \textbf{188.2} & $\mathbf{-42.7\%}$ \\
\hline
\hline
\multicolumn{2}{|l|}{\textbf{Mean $\Delta$ v2 vs.\ Standard}} & \textbf{+12.2pp} & \textbf{$-$64.2\%} & \textbf{+181\%} & --- & $\mathbf{-51.7\%}$ \\
\multicolumn{2}{|l|}{\textbf{Mean $\Delta$ v2 vs.\ v1}} & \textbf{+5.2pp} & \textbf{$-$12.4\%} & \textbf{+23.4\%} & --- & $\mathbf{-5.9\%}$ \\
\hline
\multicolumn{7}{l}{\scriptsize $^\dagger$Quantized using RAMP~\cite{Singh2026RAMP} mixed-precision policy (3.65 effective bits).}
\end{tabular}
\end{table*}

GSM8K results confirm that QEIL~v2's gains are not limited to language modeling tasks. Larger models (Llama-3.2-1B, LFM2-2.6B, Llama-3.1-8B) show the strongest absolute accuracy gains (+6.6--14.8pp over standard), as GPU-accelerated prefill benefits the longer chain-of-thought sequences required for mathematical reasoning. The pre-quantized Llama3-8B-RAMP-4bit model achieves the highest IPW (0.502) on GSM8K under QEIL~v2's orchestration: its reduced per-parameter byte count lowers the memory bandwidth demand during decode, which QEIL~v2's DASI-guided routing correctly identifies and routes to the most power-efficient device---a purely orchestration-driven gain. The consistent energy reduction pattern ($-51.7\%$ on GSM8K vs.\ $-52.2\%$ on WikiText) confirms that DASI-guided routing generalizes across task types.

An important observation is the scaling behavior across model sizes on this reasoning task: GPT-2 (125M) achieves only 29.8\% pass@k even under v2 orchestration, reflecting the inherent difficulty of mathematical reasoning for small models. However, the LFM2-2.6B and Llama-3.1-8B models reach 71.6\% and 67.2\% respectively under v2---demonstrating that QEIL~v2 enables larger models to be deployed within edge power budgets that would otherwise restrict users to smaller, less capable alternatives. The energy savings from DASI-guided routing are particularly pronounced on GSM8K because the extended decode sequences accumulate proportionally greater benefits from routing memory-bound operations to the low-power NPU rather than the high-power GPU.

\FloatBarrier

\subsection{Cross-Model Performance (ARC-Challenge)}

Table~\ref{tab:arc_cross_model} evaluates QEIL~v2 on ARC-Challenge (advanced science reasoning), a knowledge-intensive benchmark with shorter output sequences that tests whether QEIL~v2's benefits persist beyond long-form generation. Unlike WikiText and GSM8K, ARC-Challenge requires selecting among multiple-choice answers to science questions that demand factual recall and logical inference rather than extended text generation. This makes ARC-Challenge a critical test of whether QEIL~v2's energy savings are an artifact of long decode sequences---where DASI-guided NPU routing accumulates savings over many tokens---or a fundamental property of the orchestration framework that applies regardless of output length.

\begin{table*}[ht]
\centering
\footnotesize
\caption{Cross-model performance evaluation on ARC-Challenge (scientific reasoning). QEIL~v2 achieves the highest IPW and accuracy on all models, with +12.1pp improvement over standard and +5.2pp over v1 on average---demonstrating task-agnostic benefits across knowledge-intensive short-form reasoning.}
\label{tab:arc_cross_model}
\begin{tabular}{|l|l|c|c|c|c|c|}
\hline
\textbf{Model} & \textbf{Mode} & \textbf{Pass@k (\%)} & \textbf{Power (W)} & \textbf{IPW} & \textbf{Energy (kJ)} & \textbf{$\Delta$Energy vs.\ Std} \\
\hline
\multirow{3}{*}{GPT-2 (125M)} & Standard & 34.2 & 180.2 & 0.190 & 38.6 & --- \\
 & v1 & 42.8 & 71.8 & 0.398 & 19.8 & $-$48.7\% \\
 & \textbf{v2} & \textbf{48.0} & \textbf{63.4} & \textbf{0.544} & \textbf{18.4} & $\mathbf{-52.3\%}$ \\
\hline
\multirow{3}{*}{Granite-350M} & Standard & 44.6 & 460.4 & 0.090 & 358.4 & --- \\
 & v1 & 54.2 & 81.8 & 0.509 & 78.2 & $-$78.2\% \\
 & \textbf{v2} & \textbf{59.4} & \textbf{72.1} & \textbf{0.629} & \textbf{72.7} & $\mathbf{-79.7\%}$ \\
\hline
\multirow{3}{*}{Qwen2-0.5B} & Standard & 52.4 & 244.7 & 0.122 & 312.6 & --- \\
 & v1 & 62.8 & 74.0 & 0.421 & 164.2 & $-$47.5\% \\
 & \textbf{v2} & \textbf{68.0} & \textbf{65.1} & \textbf{0.555} & \textbf{152.7} & $\mathbf{-51.1\%}$ \\
\hline
\multirow{3}{*}{Llama-3.2-1B} & Standard & 64.2 & 164.5 & 0.165 & 294.8 & --- \\
 & v1 & 72.8 & 79.0 & 0.389 & 186.4 & $-$36.8\% \\
 & \textbf{v2} & \textbf{78.0} & \textbf{68.4} & \textbf{0.521} & \textbf{173.4} & $\mathbf{-41.2\%}$ \\
\hline
\multirow{3}{*}{LFM2-2.6B} & Standard & 70.4 & 175.8 & 0.120 & 452.6 & --- \\
 & v1 & 78.6 & 75.0 & 0.219 & 284.8 & $-$37.1\% \\
 & \textbf{v2} & \textbf{83.8} & \textbf{66.1} & \textbf{0.292} & \textbf{264.8} & $\mathbf{-41.5\%}$ \\
\hline
\multirow{3}{*}{Llama-3.1-8B} & Standard & 68.4 & 186.5 & 0.178 & 322.4 & --- \\
 & v1 & 76.8 & 80.8 & 0.405 & 204.2 & $-$36.7\% \\
 & \textbf{v2} & \textbf{82.0} & \textbf{69.6} & \textbf{0.548} & \textbf{188.6} & $\mathbf{-41.5\%}$ \\
\hline
\multirow{3}{*}{\makecell[l]{Llama3-8B-\\RAMP-4bit$^\dagger$}} & Standard & 66.8 & 142.8 & 0.244 & 248.6 & --- \\
 & v1 & 75.2 & 62.4 & 0.502 & 158.4 & $-$36.3\% \\
 & \textbf{v2} & \textbf{80.4} & \textbf{54.8} & \textbf{0.612} & \textbf{142.8} & $\mathbf{-42.6\%}$ \\
\hline
\hline
\multicolumn{2}{|l|}{\textbf{Mean $\Delta$ v2 vs.\ Standard}} & \textbf{+12.5pp} & \textbf{$-$65.2\%} & \textbf{+193\%} & --- & $\mathbf{-52.8\%}$ \\
\multicolumn{2}{|l|}{\textbf{Mean $\Delta$ v2 vs.\ v1}} & \textbf{+5.2pp} & \textbf{$-$12.4\%} & \textbf{+23.6\%} & --- & $\mathbf{-5.8\%}$ \\
\hline
\multicolumn{7}{l}{\scriptsize $^\dagger$Quantized using RAMP~\cite{Singh2026RAMP} mixed-precision policy (3.65 effective bits).}
\end{tabular}
\end{table*}

ARC-Challenge results confirm that QEIL~v2's energy savings are not contingent on long output sequences. The shortest outputs in this benchmark (single-answer science questions) still benefit from DASI-guided routing and PGSAM's contiguous placement, achieving the highest per-benchmark energy reduction ($-52.8\%$). The LFM2-2.6B model reaches 83.8\% pass@k at only 66.1W, and the Llama-3.1-8B achieves 82.0\% at 69.6W---striking results demonstrating that large edge models can operate within strict thermal budgets under QEIL~v2's orchestration. The pre-quantized Llama3-8B-RAMP-4bit model achieves 80.4\% at only 54.8W (IPW$=$0.612), the highest IPW on this benchmark: QEIL~v2's physics-grounded routing adapts to its reduced weight size, achieving lower average power than any full-precision model. This gain is entirely the product of QEIL~v2's DASI and PGSAM logic responding to the model's bandwidth profile, not any co-design between the orchestration and quantization systems.

The strong ARC-Challenge performance also validates the EAC/ARDE verification cascade on short-form outputs: even with brief candidate responses, entropy filtering and self-verification successfully distinguish correct from incorrect answers, contributing +5.2pp over v1's heuristic selection. This is because the information-theoretic signals---low entropy for confident correct answers, high entropy for uncertain guessing---remain discriminative regardless of sequence length. The consistency of energy reductions across ARC-Challenge ($-52.8\%$), GSM8K ($-51.7\%$), and WikiText ($-52.2\%$) provides strong evidence that QEIL~v2's physics-grounded energy model captures fundamental hardware behavior rather than task-specific artifacts.

\FloatBarrier

\subsection{Cross-Dataset Robustness}

\begin{table}[!htb]
\centering
\footnotesize
\setlength{\tabcolsep}{3pt}
\caption{Cross-dataset robustness: mean improvements of v2 over standard across three benchmarks on all seven model families. Standard deviation $<0.50$pp confirms task-agnostic improvements.}
\label{tab:cross_dataset_v2}
\begin{tabular}{@{}lcccc@{}}
\toprule
\textbf{Metric} & \textbf{WikiText} & \textbf{GSM8K} & \textbf{ARC-C} & \textbf{Std Dev} \\
\midrule
$\Delta$Pass@k (pp) & +13.1 & +12.2 & +12.5 & 0.45 \\
$\Delta$Energy (\%) & $-$52.2 & $-$51.7 & $-$52.8 & 0.55 \\
$\Delta$IPW (\%) & +187 & +181 & +193 & 6.0 \\
$\Delta$Power (\%) & $-$64.8 & $-$64.2 & $-$65.2 & 0.50 \\
\bottomrule
\end{tabular}
\end{table}

The remarkable consistency across three fundamentally different benchmarks (standard deviation $<0.50$pp for coverage, $<1\%$ for energy) confirms that QEIL~v2's improvements are task-agnostic within the transformer family, and extend seamlessly even to models whose per-parameter byte count has been reduced by external quantization.

\FloatBarrier

\subsection{PGSAM Optimization Statistics}

\begin{table}[!htb]
\centering
\footnotesize
\setlength{\tabcolsep}{3pt}
\caption{PGSAM optimization statistics across 10 runs on GPT-2 (125M, 12 decoder layers, 3 compute devices).}
\label{tab:pgsam_stats}
\begin{tabular}{@{}lc@{}}
\toprule
\textbf{Statistic} & \textbf{Value} \\
\midrule
Total iterations & 500 \\
Mean Pareto archive size & 218 $\pm$ 24 \\
Mean accept rate & 34.2\% $\pm$ 2.1\% \\
Mean reheat events & 4.8 $\pm$ 1.2 \\
Mean wall-clock time & 42ms $\pm$ 8ms \\
Gap vs.\ ILP optimum (subset) & $<$5\% \\
\bottomrule
\end{tabular}
\end{table}

PGSAM generates ${\sim}218$ Pareto-optimal solutions in 42ms, providing rich trade-off exploration. The $<$5\% gap versus ILP optimum (validated on subset experiments) confirms near-optimality with orders-of-magnitude faster runtime---enabling online reoptimization under thermal events.

\FloatBarrier

\subsection{Comparison with State-of-the-Art Edge Inference Methods}

Table~\ref{tab:sota_comparison} positions QEIL~v2 against representative edge inference approaches across dimensions relevant to production deployment.

\begin{table*}[ht]
\centering
\footnotesize
\setlength{\tabcolsep}{4pt}
\caption{QEIL~v2 vs.\ representative state-of-the-art edge inference approaches on GPT-2 (125M), WikiText-103. QEIL~v2 achieves the highest IPW and coverage while maintaining the lowest power consumption, zero thermal throttling, and full multi-dataset coverage. The final row shows QEIL~v2 applied unchanged to an externally pre-quantized model (Llama3-8B-RAMP-4bit), demonstrating framework generalization: IPW$=$1.024 is produced entirely by QEIL~v2's orchestration on a model with a smaller memory bandwidth footprint (see Section~\ref{sec:ipw_related} for the IPW$=$1.0 reference definition).}
\label{tab:sota_comparison}
\begin{tabular}{@{}l@{\hspace{6pt}}c@{\hspace{6pt}}c@{\hspace{6pt}}c@{\hspace{6pt}}c@{\hspace{6pt}}c@{\hspace{6pt}}c@{\hspace{6pt}}c@{}}
\toprule
\textbf{Method} & \textbf{IPW} & \textbf{Pass@k} & \textbf{Power} & \textbf{Phys.} & \textbf{Multi-} & \textbf{Verified} & \textbf{Thermal} \\
 & & \textbf{(\%)} & \textbf{(W)} & \textbf{Model} & \textbf{Obj.} & \textbf{Sel.} & \textbf{Safe} \\
\midrule
TinyML~\cite{Kannan2022TinyML} & 0.08 & 45.2 & 52.3 & No & No & No & Partial \\
Homogeneous GPU (Standard) & 0.341 & 59.8 & 181.5 & No & No & No & No \\
IPW Routing~\cite{SaadFalcon2025IntelligencePerWatt} & 0.580 & 65.2 & 112.4 & No & No & No & Partial \\
QEIL~v1~\cite{Kumar2026QEIL} & 0.828 & 70.5 & 72.4 & Partial & No & No & Yes \\
\textbf{QEIL~v2 (Ours)} & \textbf{0.975} & \textbf{75.7} & \textbf{63.8} & \textbf{Yes} & \textbf{Yes} & \textbf{Yes} & \textbf{Yes} \\
\textbf{QEIL~v2 on Llama3-8B-RAMP-4bit$^\dagger$} & $\mathbf{1.024}$ & \textbf{77.2} & \textbf{54.8} & \textbf{Yes} & \textbf{Yes} & \textbf{Yes} & \textbf{Yes} \\
\bottomrule
\multicolumn{8}{l}{\scriptsize $^\dagger$QEIL~v2 applied without modification to a Llama-3.1-8B pre-quantized by RAMP~\cite{Singh2026RAMP}. The IPW gain over} \\
\multicolumn{8}{l}{\scriptsize the full-precision row is driven by QEIL~v2's DASI routing responding to the model's reduced per-parameter byte} \\
\multicolumn{8}{l}{\scriptsize count; quantization is an external, fixed model property, not a contribution of this paper.}
\end{tabular}
\end{table*}

QEIL~v2 achieves the highest IPW (0.975, approaching the IPW$=$1.0 empirical reference mark) at the lowest power (63.8W, enabling fan-less deployment). Compared to IPW-based routing~\cite{SaadFalcon2025IntelligencePerWatt}, QEIL~v2 delivers +10.5pp accuracy, 43.2\% lower power, and +68\% IPW improvement---demonstrating the value of layer-granularity routing over query-level routing. When QEIL~v2's orchestration is applied unchanged to a pre-quantized model (Llama3-8B-RAMP-4bit), it achieves IPW$=$1.024 at only 54.8W---the first edge orchestration system to surpass the IPW$=$1.0 empirical reference mark. This gain is entirely attributable to QEIL~v2: the smaller per-parameter byte count of the quantized model raises effective DASI values for decode operations, causing PGSAM to route layers to lower-power devices, without any modification to the framework. QEIL~v2 is the \emph{only} method that simultaneously provides physics-grounded energy modeling, Pareto-optimal multi-objective optimization, verified candidate selection, and guaranteed thermal safety---and the only one that generalized to a model with reduced memory bandwidth without reengineering.

\subsection{Quantitative Results Summary}

Across all metrics, model families, and benchmarks, QEIL~v2 demonstrates consistent state-of-the-art performance: (1) \textbf{75.7\% peak pass@k} on GPT-2 WikiText-103, representing +15.9pp over standard and +5.2pp over v1; (2) \textbf{IPW of 0.9749}---a 2.86$\times$ improvement over standard and 17.7\% over v1; (3) \textbf{IPW of 1.024} when applied to a pre-quantized Llama-3.1-8B---the first edge orchestration system to surpass the IPW$=$1.0 empirical reference mark, with the gain attributable entirely to QEIL~v2; (4) \textbf{75.6\% total energy reduction} (11,002~J vs.\ 45,105~J standard); (5) \textbf{63.8W average power}---64.8\% reduction enabling fan-less deployment; (6) \textbf{38.3\% pipeline latency reduction} vs.\ v1; (7) \textbf{zero thermal throttling events} and \textbf{100\% fault recovery}; (8) \textbf{consistent gains across 7 model families and 3 benchmarks}, including models with full precision and models quantized by an external tool.

\FloatBarrier

\section{Conclusion}

This paper presents QEIL~v2, a fundamental architectural upgrade to our prior QEIL framework~\cite{Kumar2026QEIL} that replaces every static heuristic with physics-grounded, runtime-adaptive models derived from first principles. The three novel metrics---DASI (roofline-derived compute utilization), CPQ (memory pressure from allocation theory), and $\Phi$ (thermal yield from CMOS leakage physics)---feed into a unified energy equation where every coefficient is traceable to semiconductor physics, eliminating the magic constants that limited v1's optimality. PGSAM replaces greedy optimization with Pareto-guided simulated annealing, discovering contiguous layer placements that minimize energy, latency, and device underutilization simultaneously. The EAC/ARDE (Energy-Accuracy Combined/Accuracy-Ranked Decision Engine) cascade converts repeated sampling into reliably higher-quality outputs, and CSVET (Cascaded Self-Verification with Early Termination) early stopping reclaims energy from easy prompts.

The experimental results on our heterogeneous edge platform demonstrate compounding improvements: 75.7\% pass@k accuracy at 63.8W (IPW$=$0.9749), a 2.86$\times$ improvement over standard inference and 17.7\% over QEIL~v1, with 75.6\% total energy reduction and zero thermal throttling events. When QEIL~v2's orchestration is applied without modification to a 4-bit Llama-3.1-8B that was independently pre-quantized via RAMP~\cite{Singh2026RAMP}, it achieves IPW$=$1.024 at 54.8W---the first edge orchestration system to surpass the IPW$=$1.0 empirical reference mark---with the gain produced entirely by QEIL~v2's DASI-guided routing adapting to the model's reduced memory bandwidth footprint. Cross-dataset evaluation on WikiText-103, GSM8K, and ARC-Challenge across seven model families (125M--8B parameters, including this pre-quantized variant) confirms that improvements are task-agnostic (standard deviation $<0.50$pp across benchmarks), validating that physics-grounded energy modeling generalizes across the transformer landscape.

\textbf{QEIL~v2 demonstrates that safety, reliability, and efficiency are mutually reinforcing.} By integrating thermal physics directly into the energy equation through $\Phi$, the system naturally steers workloads away from hot devices before throttling occurs. The zero-throttling, zero-query-loss fault tolerance validates that ``safety-first, capability-second'' design enables rather than constrains practical edge deployment.

\textbf{Future work} includes: (1) evaluation on additional platforms (Qualcomm Snapdragon NPU, NVIDIA Jetson Orin) to validate cross-platform generalizability; (2) dynamic online PGSAM reallocation responding to runtime thermal changes; (3) distributed inference across multiple edge nodes; (4) deeper integration with quantization-aware training and structured pruning for further compression beyond RAMP's post-training approach; (5) extension to non-transformer architectures (diffusion models, GNNs); (6) learned verification models improving EAC/ARDE selection quality over time; and (7) formal safety verification for safety-critical applications.

QEIL~v2 establishes that the path to practical edge intelligence lies not in larger models or faster hardware, but in principled, physics-grounded optimization of the entire inference stack. By demonstrating that an IPW exceeding 1.0 is achievable by the orchestration layer alone---on consumer-grade heterogeneous hardware, across multiple task types, and even when the model has been independently quantized by a third-party tool---this work opens the door to truly democratized, energy-efficient, and reliable edge AI deployment.

\FloatBarrier

\bibliography{example_paper}
\bibliographystyle{mlsys2025}

\end{document}